\documentclass{article}

\usepackage{PRIMEarxiv}

\usepackage[utf8]{inputenc} 
\usepackage[T1]{fontenc}    
\usepackage{hyperref}       
\usepackage{url}            
\usepackage{booktabs}       
\usepackage{amsfonts}       
\usepackage{amsmath}
\usepackage{nicefrac}       
\usepackage{microtype}      
\usepackage{lipsum}
\usepackage{fancyhdr}       
\usepackage{graphicx}       
\graphicspath{{media/}}     

\title{Feasibility study and thermoeconomic analysis of cooling and heating systems using soil
for a residential and greenhouse building

}

\author{
 {Morteza Bodaghi}
  \\
   \\
  University of Louisiana at Lafayette\\
  Lafayette, USA\\
  \texttt{morteza.bodaghi1@louisiana.edu} \\
   \And
  Kazem Esmailpour, Nima Refahati \\
   \\
  University of Science and Research \\
  Tehran, Iran\\
  \texttt{esmailpour@damavandiau.ac.ir,}\\{ nimarefahati@yahoo.com} 
  {}
}

\begin{document}
\maketitle

\begin{abstract}
In the past decade, the use of renewable energy for heating and residential and greenhouse cooling structures has gained much interest due to the energy crisis, population growth, and the quantity of demand. This paper investigates heat transport and thermodynamic equations for a residential and greenhouse structure to simulate and examine the performance of a soil air conditioning system using fluid flow rate, pipe diameter, length, and fluid type features. The results show that the air-driven ventilation systems during summer outperform the rest of the HVAC systems. Moreover, decreasing the diameter and prolonging the pipeline positively affects the ventilation system's performance. In addition, the airflow rate positively correlates with our HVAC performance. We studied the performance of the air-conditioning systems in 4 cities in Iran. The Rasht's ventilation system is claimed to be the most effective for heating, and the Abadan cooling system showed outstanding results. We also compared the expenses of the soil cooling and conventional HVAC systems.
\end{abstract}


\section{Introduction}
In hot countries, the requirement for air conditioning accounts for more than half of energy usage. The need for air conditioners to maintain the thermal comfort of the environment would rise as a result of the expansion in urbanization and the rising demand for energy resources. This applies to all countries, including Iran\cite{de2021building}. In the past, this area has used techniques to maintain the thermal comfort of buildings and occupants without the use of mechanical devices. Considering that mechanical cooling and heating are becoming more expensive every day, monitoring the weather in order to offer Although it is not feasible to employ the old ways, the environmental circumstances are being monitored with more sensitivity. However, understanding how to use climate implementation methods and applying them rationally may help to reconnect man with nature\cite{ansarinasab2021life}. Among the resources available to generate optimal living circumstances are the natural powers that have been used for ventilation for both people and animals for thousands of years. During the 20th century, it was found that natural air ventilation along ventilation channels may be carried out with mechanical drives, such as fans. Facilities for air cleansing (such as air conditioning) and heat recovery were also supplied since the mechanical air conditioning system produced a steady airflow. However, despite the benefits of the mechanical ventilation system, the natural ventilation system has developed so much that the late 1990s may even be considered a modernizing era\cite{khan2008review}. In the meantime, architects and engineers in particular were interested in adopting a natural ventilation system that relied on the force of the air within the structure. They increased awareness of this area and made the use of natural ventilation systems in buildings more common. Of course, this attention came from various sources and for different reasons. Mechanical ventilation systems have evolved into intricate systems with many parts that demand both space and energy. This made the integrated usage of these technologies in structures difficult and skillful. In this instance, satisfying and balancing the ventilation system's tasks of high breathability and architectural quality are significant outcomes of the job. Compared to the building structure, the service life of a mechanical ventilation system is brief. On the other hand, owing to the existence of ducts that are anchored to the building's structure, the reconstruction or repair of the mechanical ventilation system shortens the building's lifespan. As a consequence, a significant portion of the construction expenditures go toward the mechanical ventilation system\cite{bharadwaj1981temperature}. The sick building phenomenon, which is a term used in studies to describe how many mechanical ventilation systems fail to provide the necessary air, has diminished confidence in mechanical systems as the best option. Both inside and outside, mechanical ventilation systems make a lot of noise, and they may be challenging to clean, repair, and maintain. All of these factors, along with growing knowledge of the negative environmental effects of using more energy and resources, have increased the emphasis on low-energy buildings\cite{inproceedings}. It is challenging to perform heat energy recovery when using natural ventilation, which may also have an impact on interior temperature changes and air quality. However, accurate management and forecasting of airflow in natural ventilation systems are now achievable because of advancements in computer technology\cite{t2012comparison}.

\subsection{The geometry of the building and its ventilation system}

In this project, the structure is envisioned as shown in Figure 1. This structure is a greenhouse that is 6 meters long and 4 meters broad for the purposes of verification; upon verification, it is regarded as a typical structure. Outside of this structure, pipes for the production system for heating and cooling may be laid at various depths and in a variety of sizes. The pipes in these systems are typically buried at a depth of one meter, horizontally\cite{gehlin2014thermo}. In this study, pipes with dimensions of 2 to 8 cm, 30 to 60 meters in length, and 1000 to 2000 meters per hour have been examined.

\begin{figure}[h!]
    \centering   \includegraphics[width=0.55\textwidth]{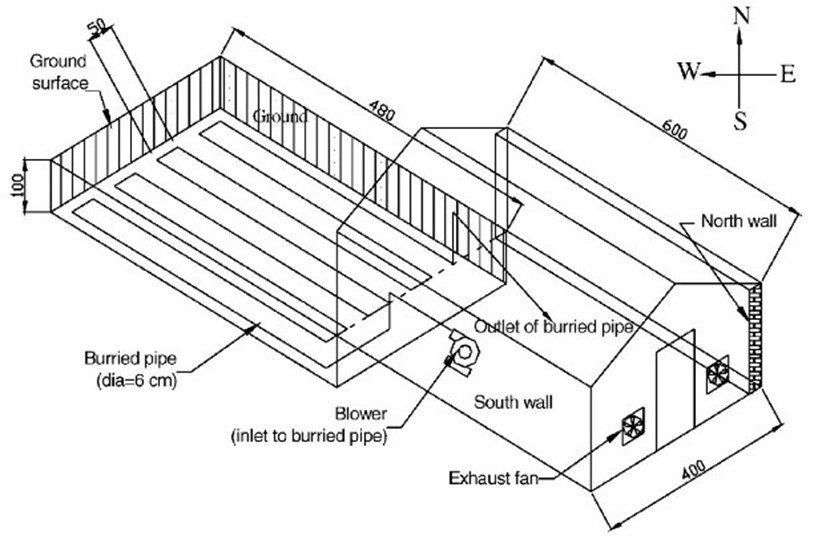}
    \caption{The investigated building \cite{bahadori1994viability}}
\end{figure}

\subsection{Governing equations}

The governing equations of this building are the energy equations for each of its parts, which are written as the following equations\cite{d2020energy}. Figure 2 indicates the energy parameters in the buried pipe, and Table 1 shows the parameters in the present study.

\begin{figure}[h!]
    \centering   \includegraphics[width=0.5\textwidth]{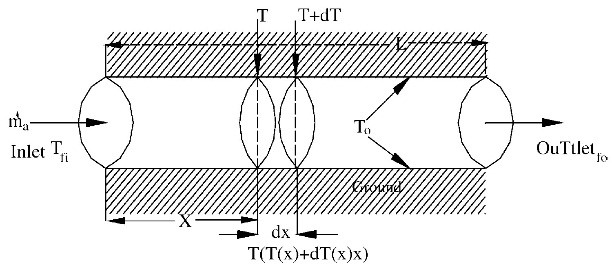}
    \caption{Energy in buried pipes for ventilation \cite{costa2006thermodynamic}}
\end{figure}

Equation 1 illustrates how the rate of energy absorbed by the north wall may be represented as being equal to the rate of energy provided from the north wall to the air within the greenhouse plus the rate of energy supplied from the north wall to the surrounding environment\cite{kusuda1965earth}.

\begin{equation}
    \alpha_{n} (1-r_n)F_n(\Sigma A_iI_i \tau_i) = h_{c.nr} (T_n - T_r) A_n + h_{nb} (T_n - T_a) A_n
\end{equation}

Equation 2 illustrates how the rate of energy absorbed by the floor may be stated as equal to the rate of energy provided from the floor to the air within the greenhouse plus the rate of energy delivered from the floor to the soil for the energy balance of the building's floor\cite{koohi2013review}.

\begin{equation}
    \alpha_f (1-r_f )(1-f_p )(1-r_n )(1-F_n )(\Sigma A_i I_i \tau_i ))=h_{c.fr} (T_f-T_r ) A_f+h_\infty (T_f-T_\infty ) A_f
\end{equation}

The rate of energy given from the floor to the air inside the greenhouse, the rate of energy given from the north wall to the air inside the greenhouse, and the rate of energy given from the considered cooling or heating system are equal to the rate of energy lost from the greenhouse to the surrounding environment, plus the rate of energy lost from the greenhouse to the surrounding environment with force.

\begin{multline}
       (1-a_n )(1-r_n ) F_n (\Sigma A_i I_i \tau_i ))+(1-a_f )(1-r_f )×(1-F_P )(1-r_n )(1-F_n )(\Sigma A_i I_i \tau_i )) h_{c.nr}\\
       (T_n-T_r ) A_n+h_{c,fr} (T_f-T_r ) A_f+\dot{Q}_u=0.33NV(T_r-T_a )+(\Sigma A_i U_i ))(T_r-T_a )+M_a C_a  (dT_r)/(dt)
       \end{multline}

\begin{table}
\centering
\caption{Introduction of the parameters in the equations}
\begin{tabular}{|| l  l ||} 
 \hline
    The area of the south wall & $A_e(m^2)$ \\
    \hline
    floor area & $A_f(m^2)$ \\
    \hline
    The area of the northern wall  & $A_n(m^2)$ \\
    \hline
    North roof area  & $A_{nr}(m^2)$ \\
    \hline
    South roof area  &$A_{sr}(m^2)$ \\
    \hline
    The area of the western wall  & $A_{ww}(m^2)$ \\
    \hline
    Heat capacity of air & $C_a(J/kg\circ C)$ \\
    \hline
    Solar energy absorption coefficient of the north wall  & $F_n$ \\
    \hline
    Solar energy absorption coefficient for plants  & $F_P$ \\
    \hline
    Heat transfer coefficient of ground to air inside the pipe  & $F_R$ \\
    \hline
    Air heat transfer coefficient of the buried pipe  & $h(W/m^2\circ C)$ \\
    \hline
    Heat transfer coefficient from the ceiling into the room  & $h_i(W/m^2\circ C)$ \\
    \hline
 
    Changed room's air per hour  & $N$\\
    \hline
    pipe radius  & $r_1(m)$ \\
    \hline
    
    Absorption coefficient of the north wall  & $\alpha_n$ \\
    \hline
    Radiation transfer  & $\tau$ \\
    \hline
    Reflectance coefficient of the north wall  & $r_n$ \\
    \hline
    Floor absorption coefficient  & $\alpha_f$ \\
    \hline
    Room size  & $V(m^3)$ \\
    \hline
    Reflectance coefficient of the room floor  & $r_f$\\
    \hline
    Total heat transfer for the room  & $U(W/m^2\circ C)$ \\
    \hline
    Speed inside the pipe  & $v(m/s^2)$ \\
    \hline
    Air amount  & $\dot{m}_a(kg/s)$ \\
    \hline
    The weight of the air in the room  & $M_a(kg)$ \\
    \hline
    length of the pipe  & $L(m)$ \\
    \hline
    Buried depth  & $L_g(m)$ \\
    \hline
    Thermal conductivity coefficient of northern brick & $K_B(W/m^{2\circ} C)$  \\
    \hline
    Earth's thermal conductivity coefficient & $K_g(W/m^{2\circ} C)$ \\
    \hline
    Heat transfer coefficient from the floor to the air inside the room  & $h_{c,fr}(W/m^2\circ C)$ \\
    \hline
     Heat transfer coefficient from the floor to the north brick wall  & $h_{c,nr}(W/m^2\circ C)$ \\
    \hline
    Heat transfer coefficient from the roof to the open air & $h_0(W/m^2\circ C)$  \\
    \hline
    Heat transfer coefficient from the north brick wall to the open air  & $h_{nb}(W/m^2\circ C)$ \\
    \hline

\end{tabular}
\end{table}

In equation (3), $\dot{Q}_u$, which is the energy rate obtained from the heating and cooling system, is equal to\cite{zhen2007district}:

\begin{align}
       {\dot{Q}}_u=F_R\dot{m}_aC_a(T_0-T_fi)
\end{align}

\begin{align}
F_R = 1 - e^{-(\frac{2\pi r_1 h}{\dot{m}_a C_a})L}
\end{align}

In the above relationship, $F_R$ is obtained as follows:

\begin{align}
       {\dot{m}}_aC_a\frac{dT(x)}{dx}=(2{\pi r}_1)h(T_0-T(x))dx
\end{align}

\begin{align}
       {\dot{m}}_a={\pi r}_1^2\rho_au
              \end{align}

\begin{align}
       h=2.8+3.0u
              \end{align}

\begin{align}
       \frac{dT(x)}{T_0-T(x)}=\frac{2{\pi r}_1h}{{\dot{m}}_aC_a}dx
              \end{align}

\begin{align}
       -Ln(T_0-T(x))=\frac{2{\pi r}_1h}{{\dot{m}}_aC_a}x+c
              \end{align}

\begin{align}
       x=0\rightarrow T(x)=T_{fi} \xrightarrow{} c=-log{(T_0-T_{fi})}
\end{align}
\begin{align}
       \frac{T(x)-T_0}{T_{fi}-T_0}=e^{-\frac{2{\pi r}_{1h}}{{\dot{m}}_aC_a}x}
              \end{align}

\begin{align}
       T(x)=T_0(1-e^{-\frac{2{\pi r}_{1h}}{{\dot{m}}_aC_a}x})+T_{fi}e^{-\frac{2{\pi r}_{1h}}{{\dot{m}}_aC_a}x}
              \end{align}

\begin{align}
       T_{f0}=T_0(1-e^{-\frac{2\pi r_{1h}}{{\dot{m}}_aC_a} L})+T_{fi}e^{-\frac{2{\pi hr}_1}{{\dot{m}}_aC_a}L}
              \end{align}

\begin{align}
       T_{f0}-T_{fi}=(T_0-T_{fi})(1-e^{-\frac{2{\pi r}_{1h}}{{\dot{m}}_aC_a}L})
              \end{align}

\begin{align}
       {\dot{Q}}_u={\dot{m}}_aC_a(T_{f0}-T_{fi})
              \end{align}

\begin{align}
       {\dot{Q}}_u={\dot{m}}_aC_a(T_0-T_{fi})(1-e^{-\frac{2{\pi r}_1h}{{\dot{m}}_aC_a}L})
              \end{align}

\begin{align}
       {\dot{Q}}_u=F_R{\dot{m}}_aC_a(T_0-T_{fi})
              \end{align}

\begin{align}
       F_R=(1-e^{-\frac{2{\pi r}_1h}{{\dot{m}}_aC_a}L})
              \end{align}
After simplifying equation 1, we get the following equation: 
\begin{align}
      h_{c,fr}(T_f-T_r)=F_2\frac{I_{effF}}{A_f}-U_f(T_r-T_\infty)
              \end{align}
which in the above equation: 
\begin{align}
      I_{effF}=\alpha_f(1-r_f)(1-F_p)(1-r_n)(1-F_n)(\sum{A_iI_i\tau_i})
              \end{align}

\begin{align}
      F_2=\frac{h_{c,fr}}{(h_{c,fr}+h_\infty)}
              \end{align}             

 \begin{align}
      U_f=\frac{(h_{c,fr})(h_\infty)}{(h_{c,fr}+h_\infty)}
              \end{align}   

\begin{align}
      T_\infty\approx T_a
              \end{align}             
\begin{align}
      T_{fi}=T_r
              \end{align}             

Equation 26, which is the last equation to be solved, is produced by combining the simplified versions of the aforementioned equations in Equation 3. This equation is used to determine the temperature within the structure. Equation 26's parameters a and F(t) come from equations 27 and 29 through 35, respectively. Equation 28 is used to determine the value of a1 in equation 27\cite{zhao2004simulation}. 

\begin{align}
      \frac{dT_r}{dt}+aT_r=F(t)
              \end{align}             
\begin{align}
      a=\frac{a_1}{M_aC_a}
              \end{align}
\begin{align}
      a_1=U_nA_n+U_fA_f+0.33NV\ +(\sum{A_iU_i})+F_R{\dot{m}}_aC_aT_0
              \end{align}
\begin{align}
      F(t)=I_{effR}+F_1I_{effN}+F_2I_{effF}+F_R{\dot{m}}_aC_aT_0
              \end{align}
\begin{align}
      (UA)_{eff}=U_nA_n+U_fA_f+0.33NV+(\sum{A_iU_i})
              \end{align}

\begin{align}
      I_{effR}=(1-a_n)(1-r_n)F_n(\sum{A_iI_i\tau_i})+(1-a_f)(1-r_f)\times(1-F_p)(1-r_n)(1-F_n)(\sum{A_iI_i\tau_i})
              \end{align}

\begin{align}
      (\sum{A_iI_i\tau_i})=A_eI_e\tau_e+A_{ww}I_{ww}\tau_{ww}\ +A_{sr}I_{sr}\tau_{sr}\ +\ A_{nr}I_{nr}\tau_{nr}\ +A_sI_s\tau_s
              \end{align}

\begin{align}
      (\sum{A_iU_i})=A_e\ U_e\ +A_{ww}U_{ww}\ +A_{sr}u_{sr}\ +A_{nr}U_{nr}+\ A_sU_S
              \end{align}

\begin{align}
      U_e=U_{ww\ =\ }U_{sr}=U_{nr\ }=U_s=U
              \end{align}

\begin{align}
      \tau_e=\tau_{ww\ }=\tau_{sr\ }=\tau_{nr\ }=\tau_s
              \end{align}
Equation 36 can be used to calculate the heat transfer coefficient between the air inside buried pipes and the soil, where $K$ is the fluid's thermal conductivity coefficient (air or water), $D$ is the pipe's diameter, ${Re}_D$ is the pipe's Reynolds number, $Pr$ is the fluid's Prandtl number, and $a$, $m$, and $n$ are other constants with varying values depending on whether the flow is turbulent or smooth\cite{sharan2003performance}. 
\begin{align}
      h_1=\pi a\frac{K}{D}{Re}_D^m{Pr}^n
              \end{align}

\section{Verification of the results}

To confirm the results as noted, a greenhouse was modeled in the form of Figure 1, which was done in the article of Mr. Qosal et al., and the results were provided as follows:

\begin{figure}[!h]
    \begin{minipage}[c]{0.46\textwidth}
         \centering
    \includegraphics[width=\textwidth]{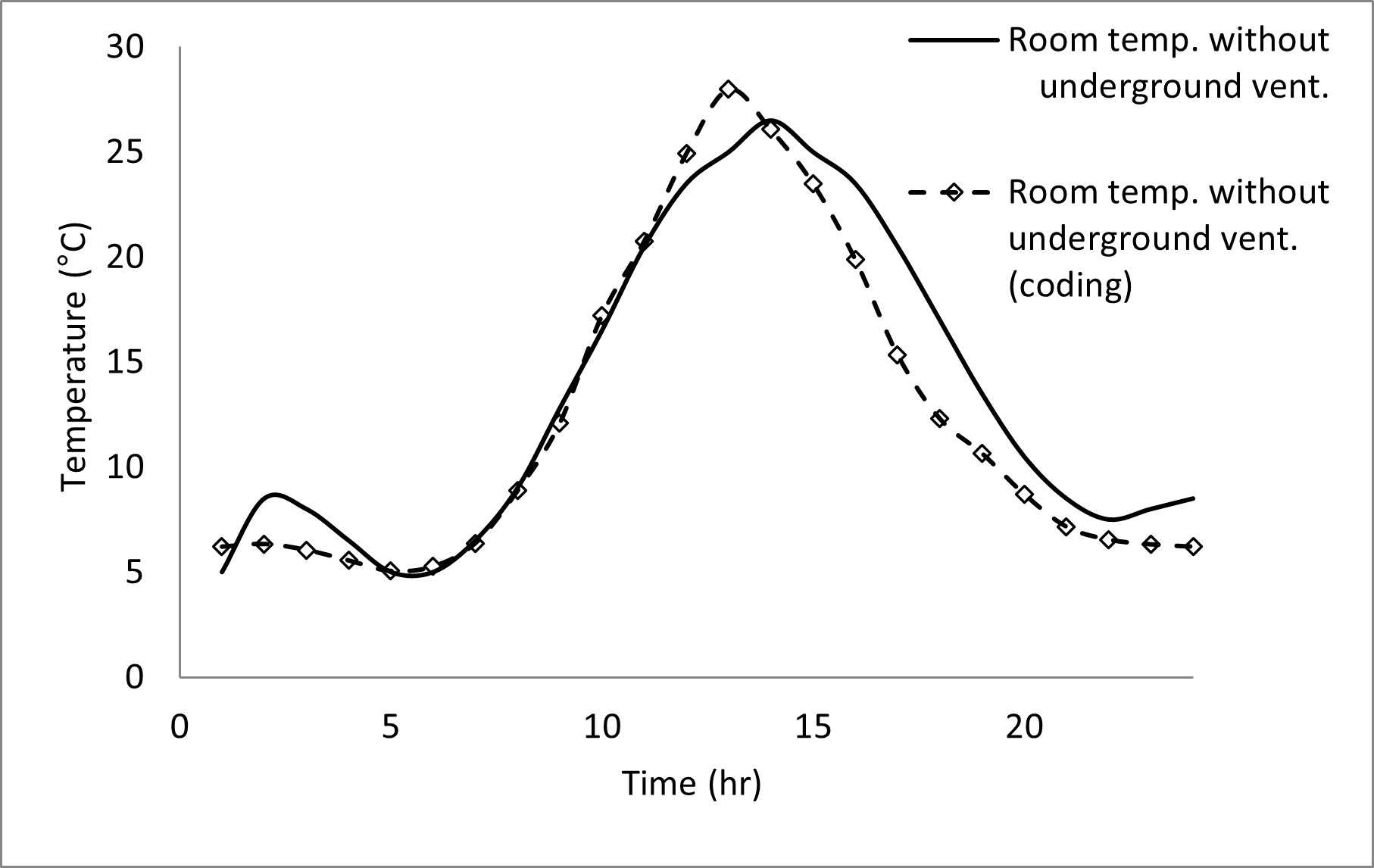}
     \caption{Comparison chart of building temperature changes without heating system for experimental data \cite{ghosal2006modeling} and coding in 24 hours}
  \end{minipage}
  \hfill
  \begin{minipage}[c]{0.46\textwidth}
     \centering
  \includegraphics[width=\textwidth]{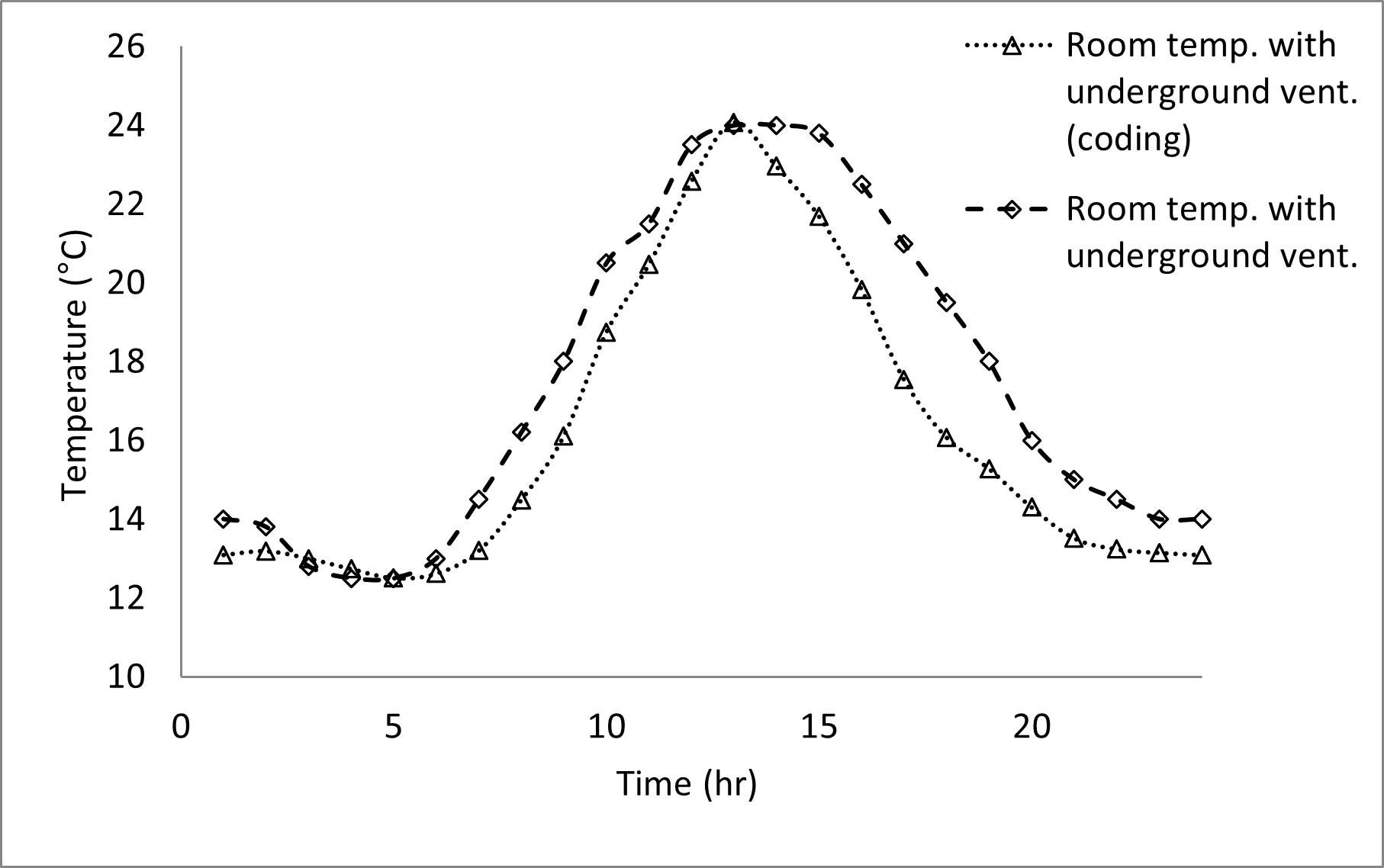}
    \caption{Comparison diagram of building temperature changes with heating system for experimental data \cite{ghosal2006modeling} and coding in 24 hours}
  \end{minipage}
\end{figure}

The practical findings from Mr. Qosal's work \cite{ghosal2006modeling} are compared to the results obtained by modeling using MATLAB software in Figures 3 and 4. As observed in Figure 3, the findings from the modeling without a heating system are almost identical to the results from Mr. Qosal's paper. Up until 12 o'clock, the graphs show the least inaccuracy, and after that, there is a difference. These two graphs both go up a little. In general, it can be claimed that the graphs in Figure 3 correspond well and support the accuracy of the findings. Figure 4, which, like Figure 3, indicates that two graphs up to 12 o'clock exhibit less inaccuracy than 12 o'clock and later, compares the data from the article with the modeling results for the heating system. He said that the graphs well matched and validated the accuracy of the findings.

\section{Results}
This section examines this heating and cooling system in four distinct cities: Tehran, Tabriz, Abadan, and Rasht. The findings for these cities for the hottest and coldest days of the year (air temperature for these cities' hottest and coldest days) are presented. The meteorological website yielded the 2015 warmest day) \cite{WinNT} is being looked at for residential and greenhouse structures. After that, it will be determined how these factors affect the heating and cooling system by choosing the parameters that have an impact on it, such as the length, diameter, and flow rate, and comparing the results. The results for the residential structure are as follows:

\begin{figure}[!h]
    \begin{minipage}[c]{0.46\textwidth}
         \centering
    \includegraphics[width=\textwidth]{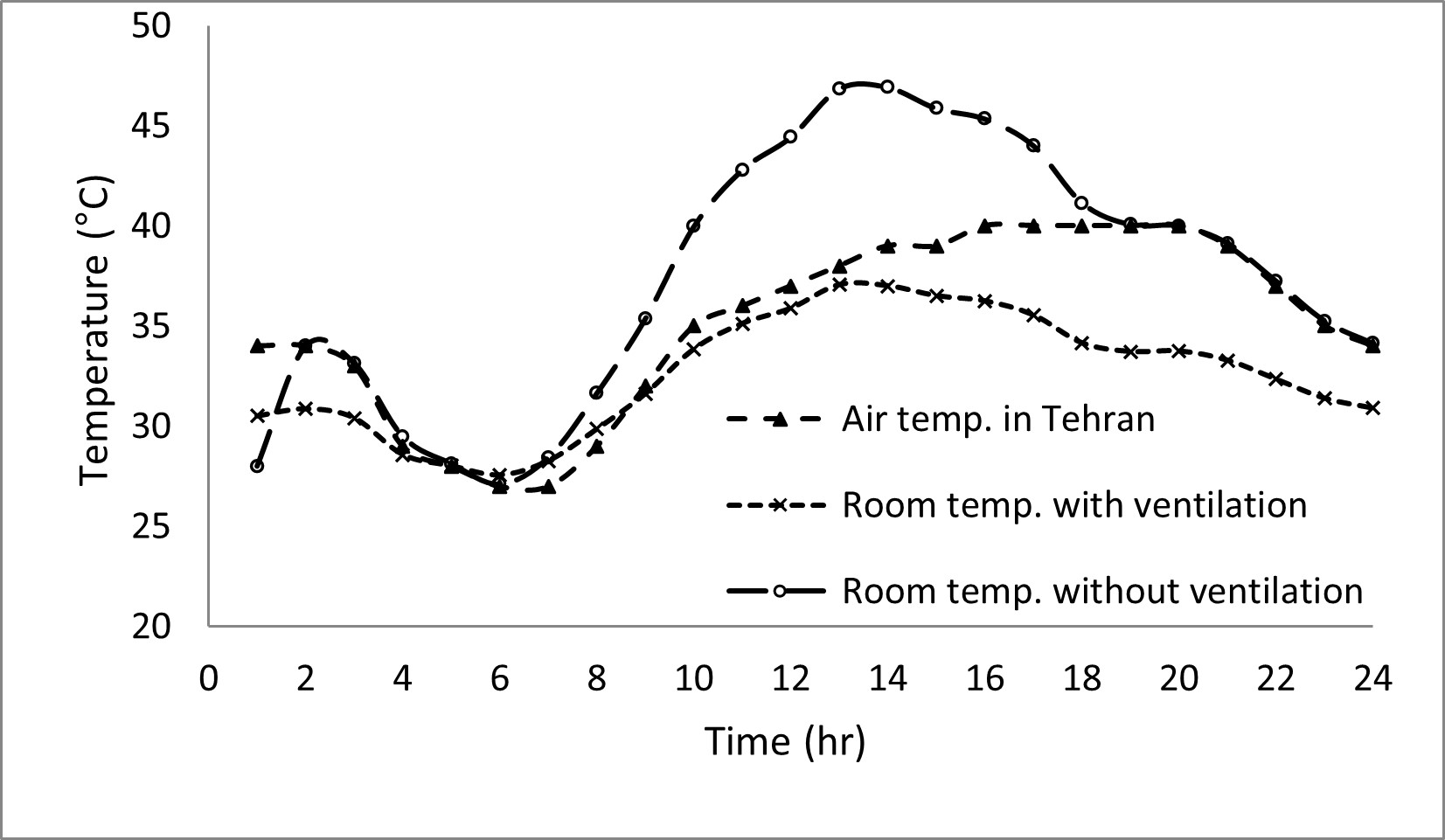}
     \caption{Temperature changes diagram at different times of the hottest day in Tehran}
  \end{minipage}
  \hfill
  \begin{minipage}[c]{0.46\textwidth}
     \centering
  \includegraphics[width=\textwidth]{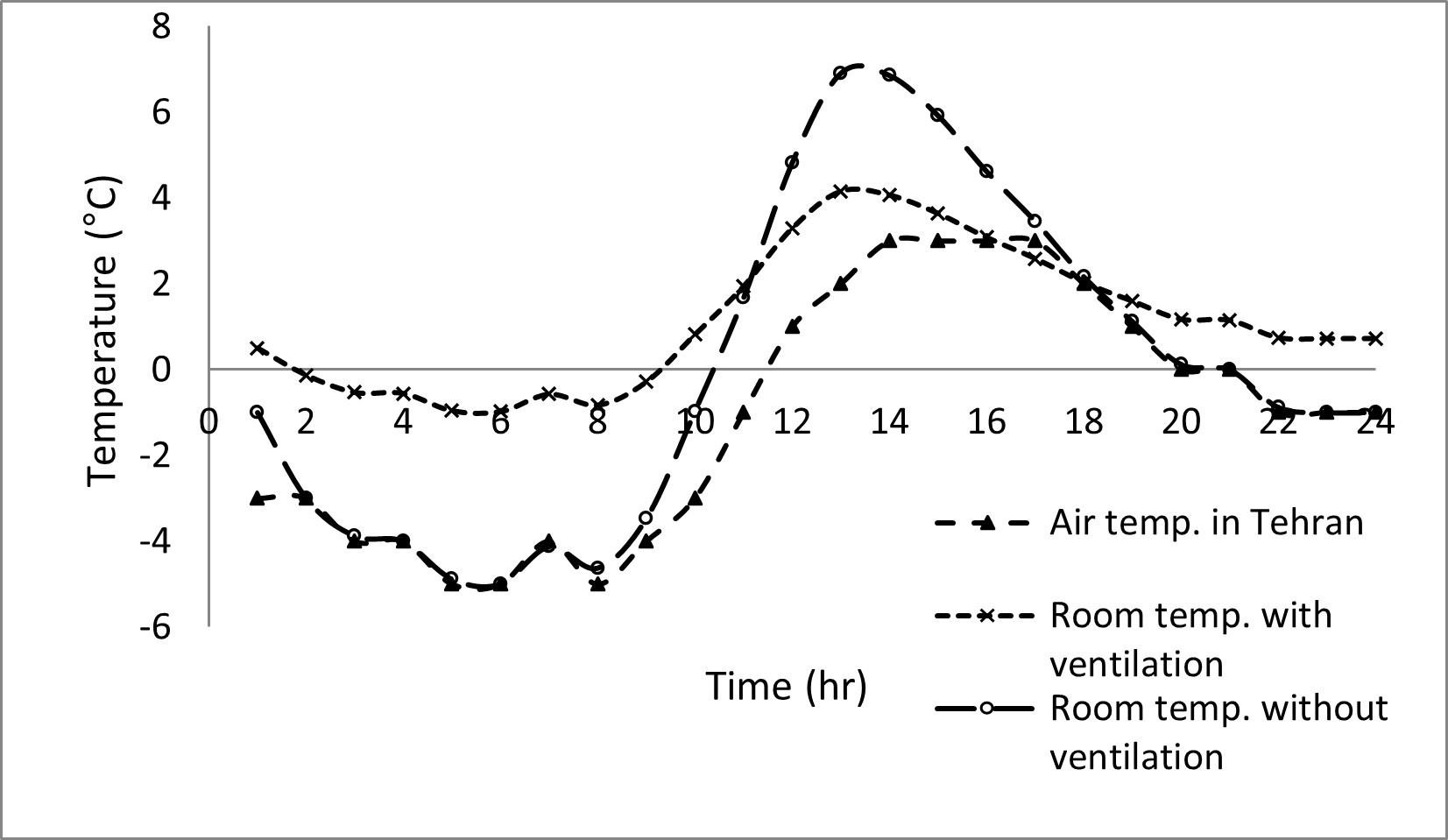}
    \caption{Temperature changes chart at different times of the coldest day in Tehran}
  \end{minipage}
\end{figure}

\begin{figure}[!h]
    \begin{minipage}[c]{0.46\textwidth}
         \centering
    \includegraphics[width=\textwidth]{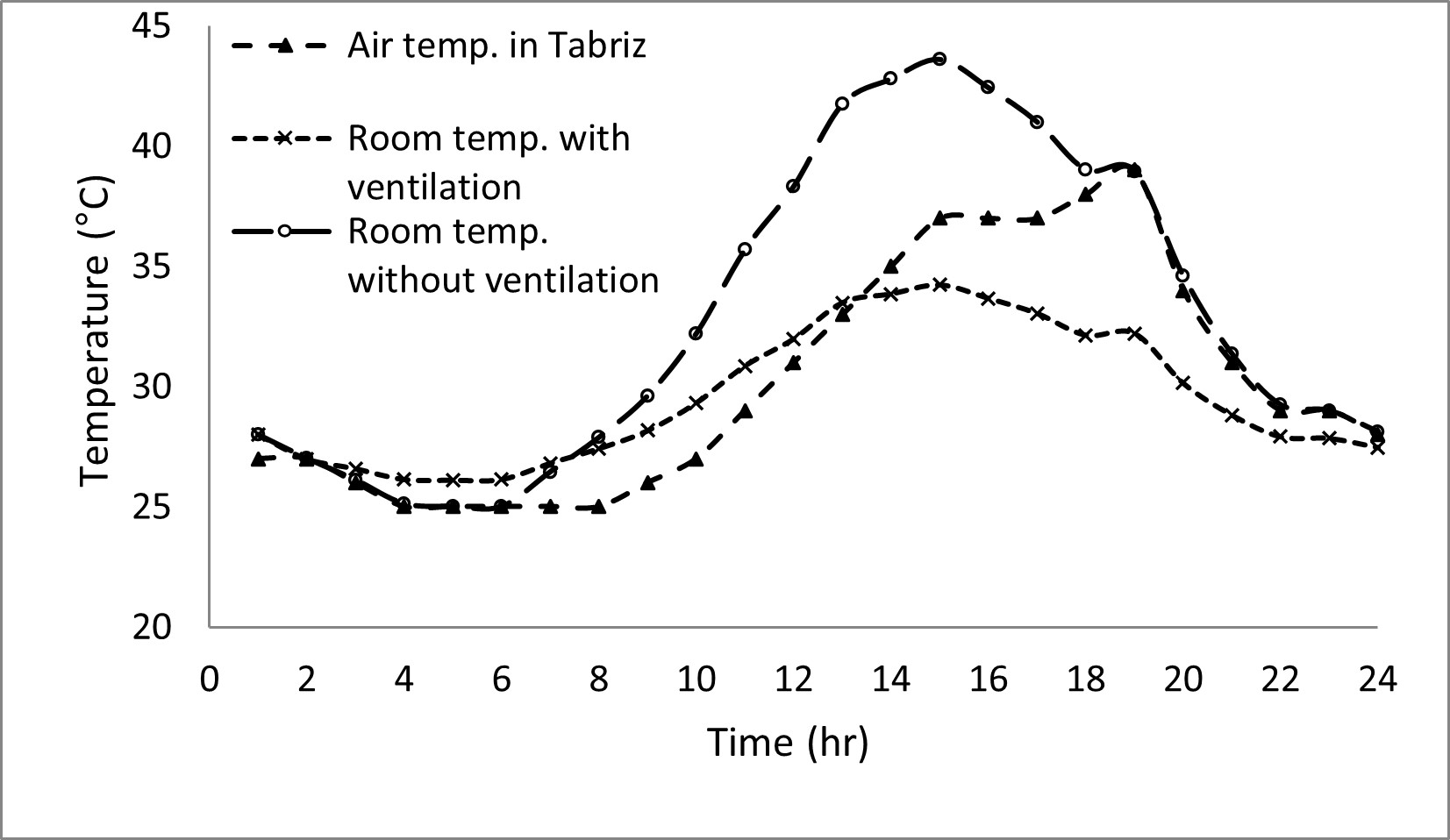}
     \caption{Temperature changes diagram at different times of the hottest day in Tabriz}
  \end{minipage}
  \hfill
  \begin{minipage}[c]{0.46\textwidth}
     \centering
  \includegraphics[width=\textwidth]{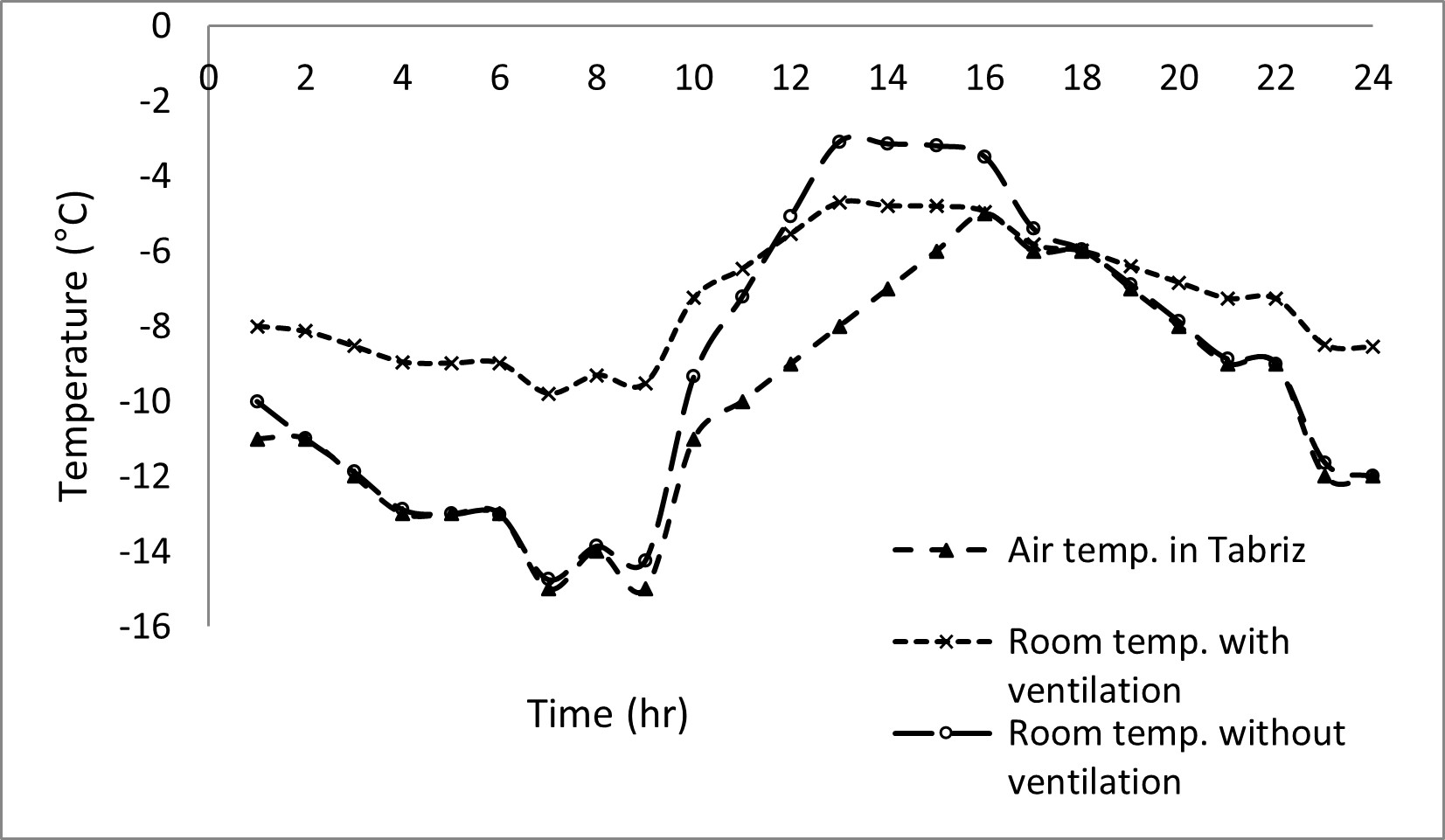}
    \caption{Temperature changes chart at different times of the coldest day in Tabriz}
  \end{minipage}
\end{figure}

\begin{figure}[!h]
    \begin{minipage}[c]{0.46\textwidth}
         \centering
    \includegraphics[width=\textwidth]{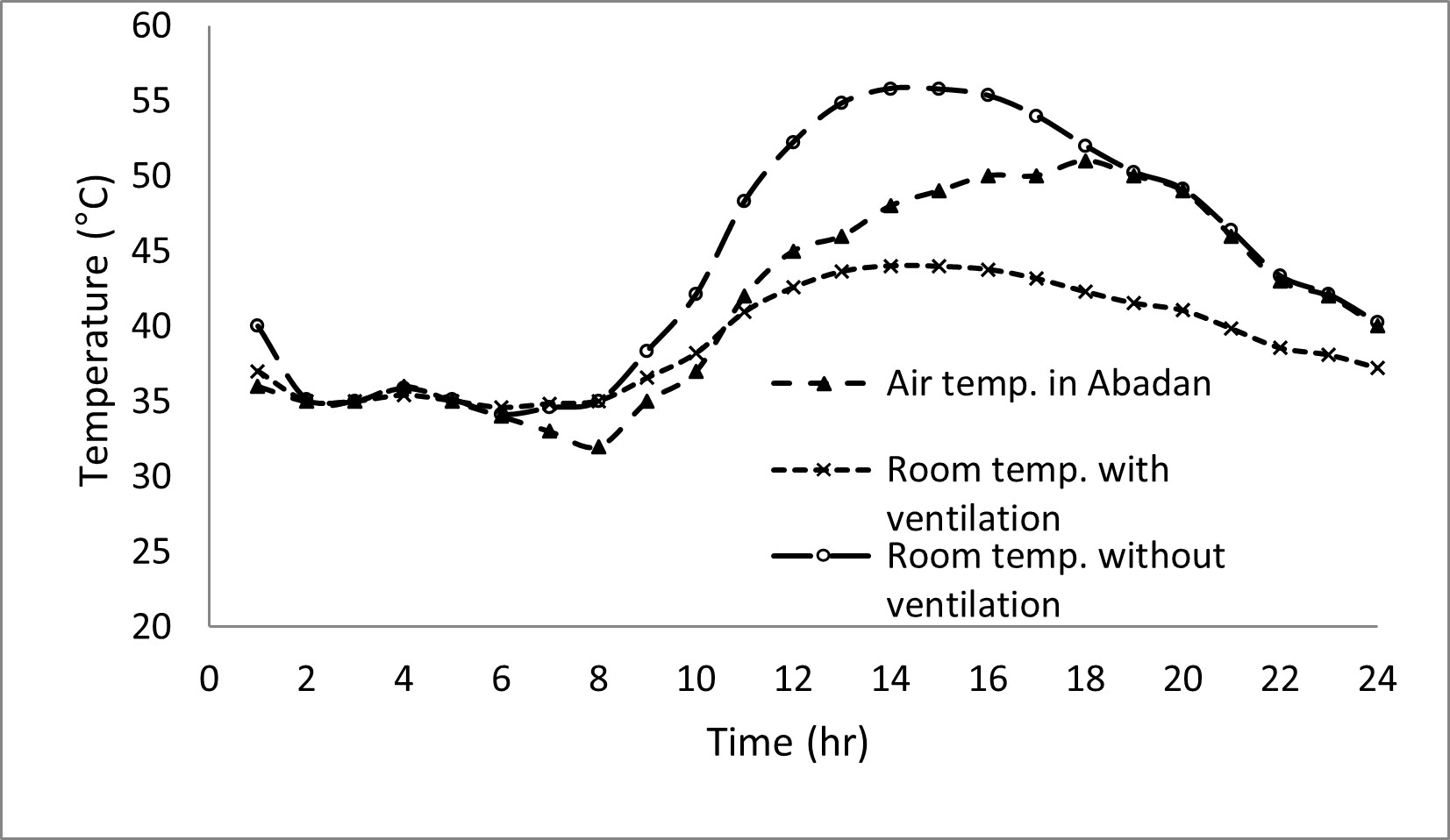}
     \caption{Temperature changes diagram at different times of the hottest day in Abadan}
  \end{minipage}
  \hfill
  \begin{minipage}[c]{0.46\textwidth}
     \centering
  \includegraphics[width=\textwidth]{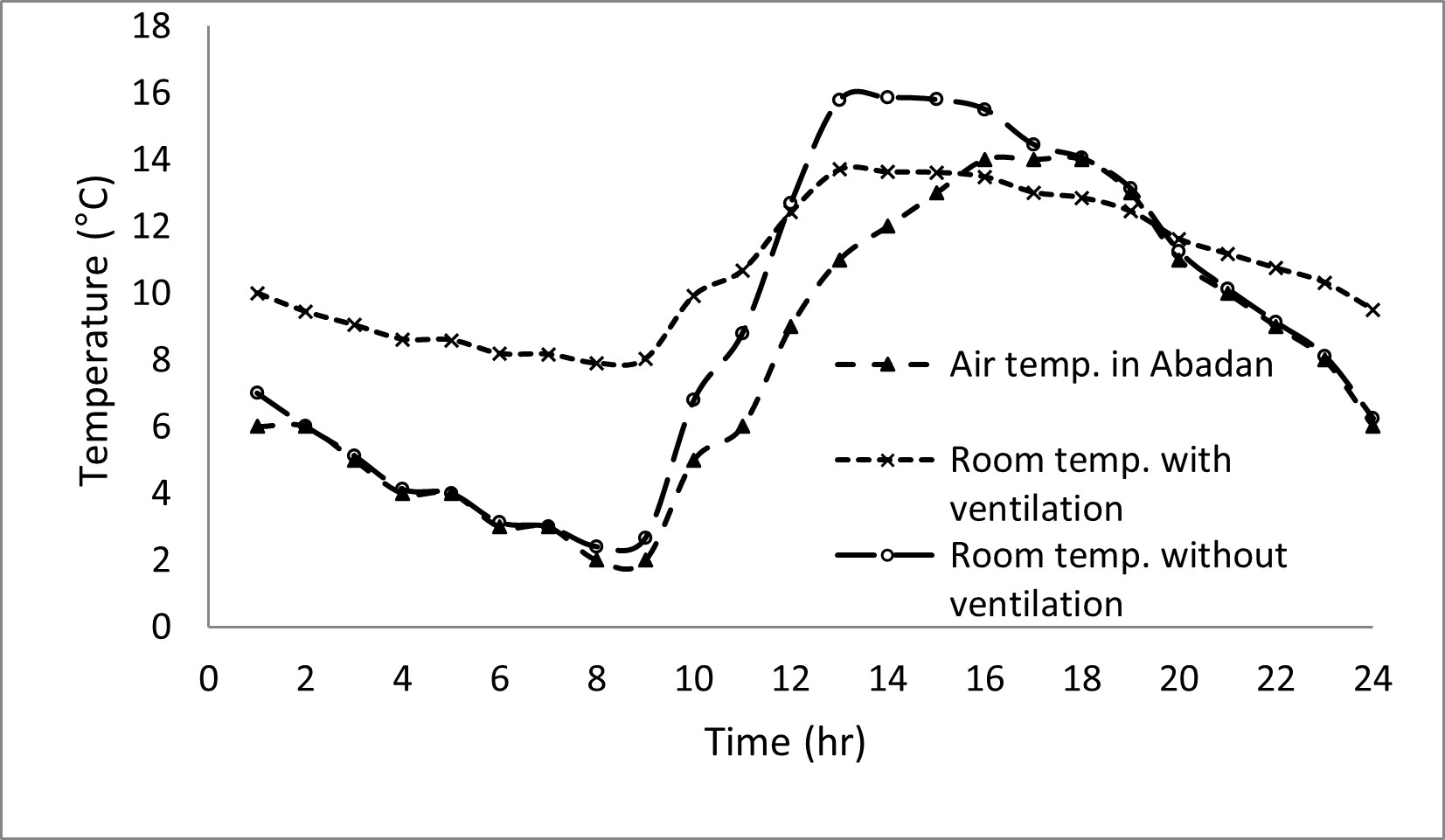}
    \caption{Temperature changes chart at different times of the coldest day in Abadan}
  \end{minipage}
\end{figure}

\begin{figure}[!h]
    \begin{minipage}[c]{0.46\textwidth}
         \centering
    \includegraphics[width=\textwidth]{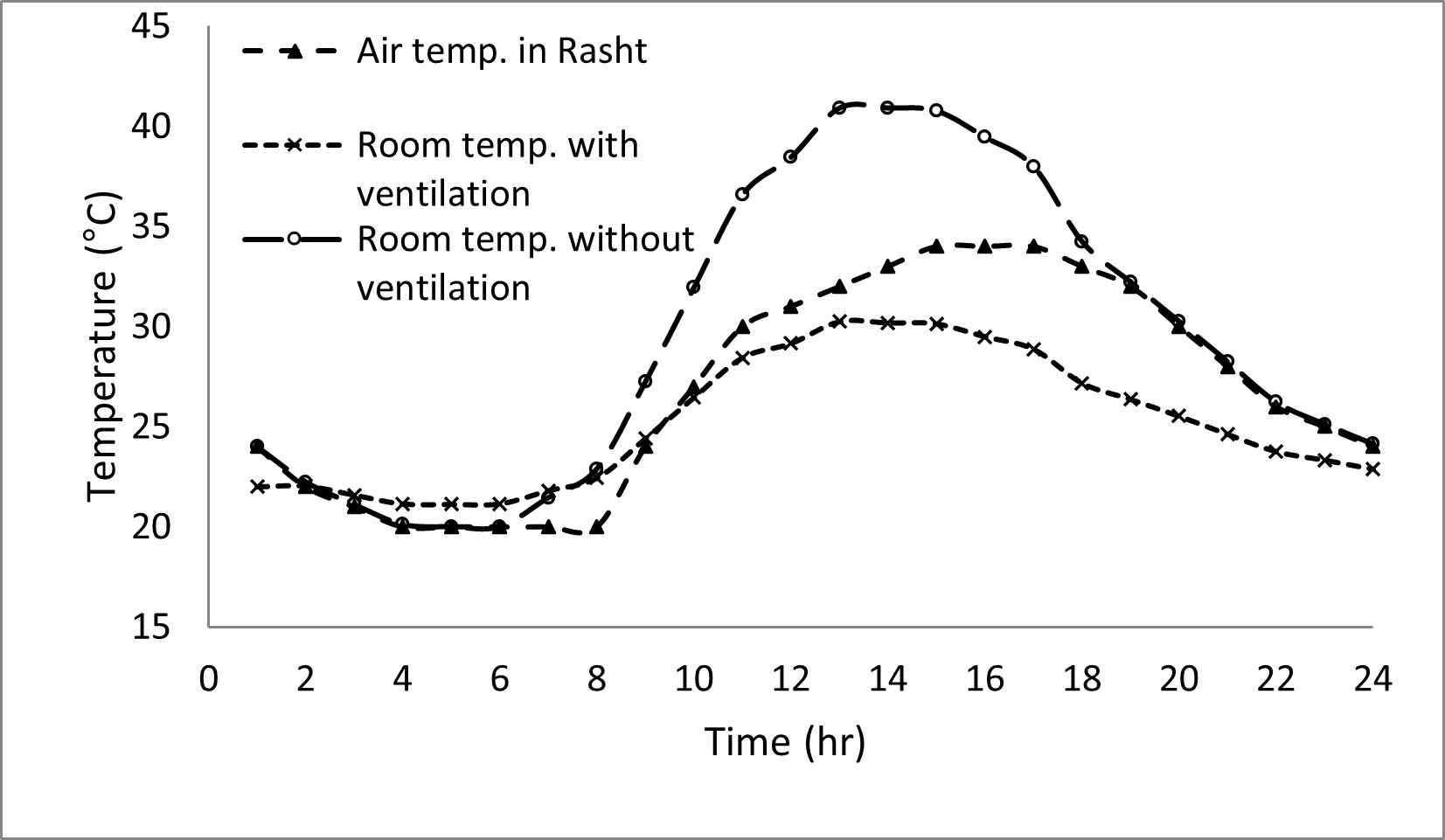}
     \caption{Temperature changes diagram at different times of the hottest day in Rasht}
  \end{minipage}
  \hfill
  \begin{minipage}[c]{0.46\textwidth}
     \centering
  \includegraphics[width=\textwidth]{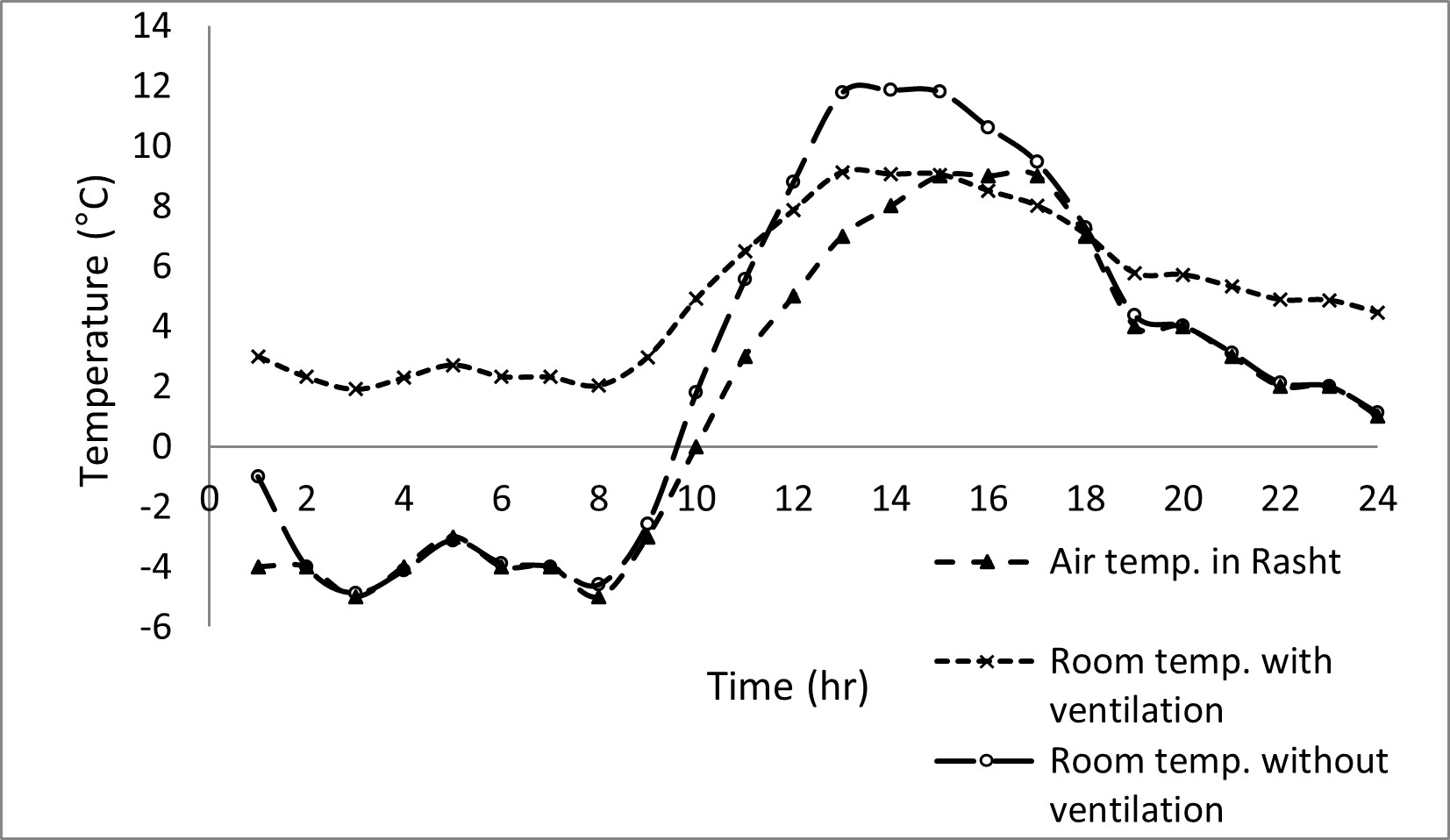}
    \caption{Temperature changes chart at different times of the coldest day in Rasht}
  \end{minipage}
\end{figure}

Figure 5 shows how the air temperature in Tehran city varied from 1 to 8 o'clock. It dropped from -3 °C with a little slope to -5 °C, then rose with dawn until it reached a maximum of -6 °C. The air temperature is at its maximum at about 13:00 with a temperature of 7 degrees, then it drops after that and almost stays constant until 20:00 and dusk. The air temperature and room temperature are quite comparable and near to one another during the hour when there is no sunshine, as shown by the indoor temperature graph without a heating system, but the more radiation present, the more apart these two graphs are from one another. It may last until 13:00 when they start to diverge the most. The temperature diagram of the heated building also demonstrates how the heating system has reduced the space between the lowest and highest temperatures, or in other words, how the lowest temperature at night has grown, and the maximum temperature has decreased. The daytime temperature has dropped as a result. In summary, it can be claimed that Tehran's heating system made the city's lowest temperature rise from -5°C to -1°C on the coldest day of the year. The lowest temperature in Tabriz City rose from -15°C to -9°C thanks to the heating system, as shown in Figure 7. Figure 9 shows how the heating system for Abadan City's coldest day increased the lowest temperature from 2°C to 8°C. Figure 11 shows how the heating system for Rasht City's coldest day raised the city's lowest temperature from -5°C to 2°C. 
As seen in Figure 6, on the warmest day in Tehran, the air temperature gradually drops from 1 to 8 o'clock, then rises to 40 degrees Celsius until 2 o'clock, where it then almost stays constant till. This temperature drops as the night draws to a close. The room temperature and the air temperature are quite comparable and near to one another during the hour when there is no sunshine, as shown by the indoor temperature graph without a cooling system, but the more radiation present, the more apart these two graphs are from one another. It may last until 13:00 when they start to diverge the most. The cooling system has also resulted in a shorter difference between the lowest temperature and the maximum temperature, as seen in the temperature diagram of the building with the cooling system. In conclusion, it can be claimed that Tehran's cooling system was responsible for the highest temperature reduction from 47 degrees Celsius to 37 degrees on the warmest day of the year. Figure 8 illustrates how Tabriz's cooling system helped the city's warmest day's maximum temperature decrease from 44°C to 34°C. Figure 10 shows how the cooling system worked to lower the highest temperature in Abadan City from 56°C to 44°C on the warmest day of the year. Figure 12 shows how Rasht City's cooling system reduces the highest temperature from 41°C to 30°C on its warmest day. 

The findings from the comparison of various ventilation system characteristics, including diameter, pipe length, intake flow, and fluid type, will be addressed in the paragraphs that follow. The pipe diameter parameter for the warmest day and the city of Tehran is first studied. 

\begin{figure}[!h]
    \centering   \includegraphics[width=0.55\textwidth]{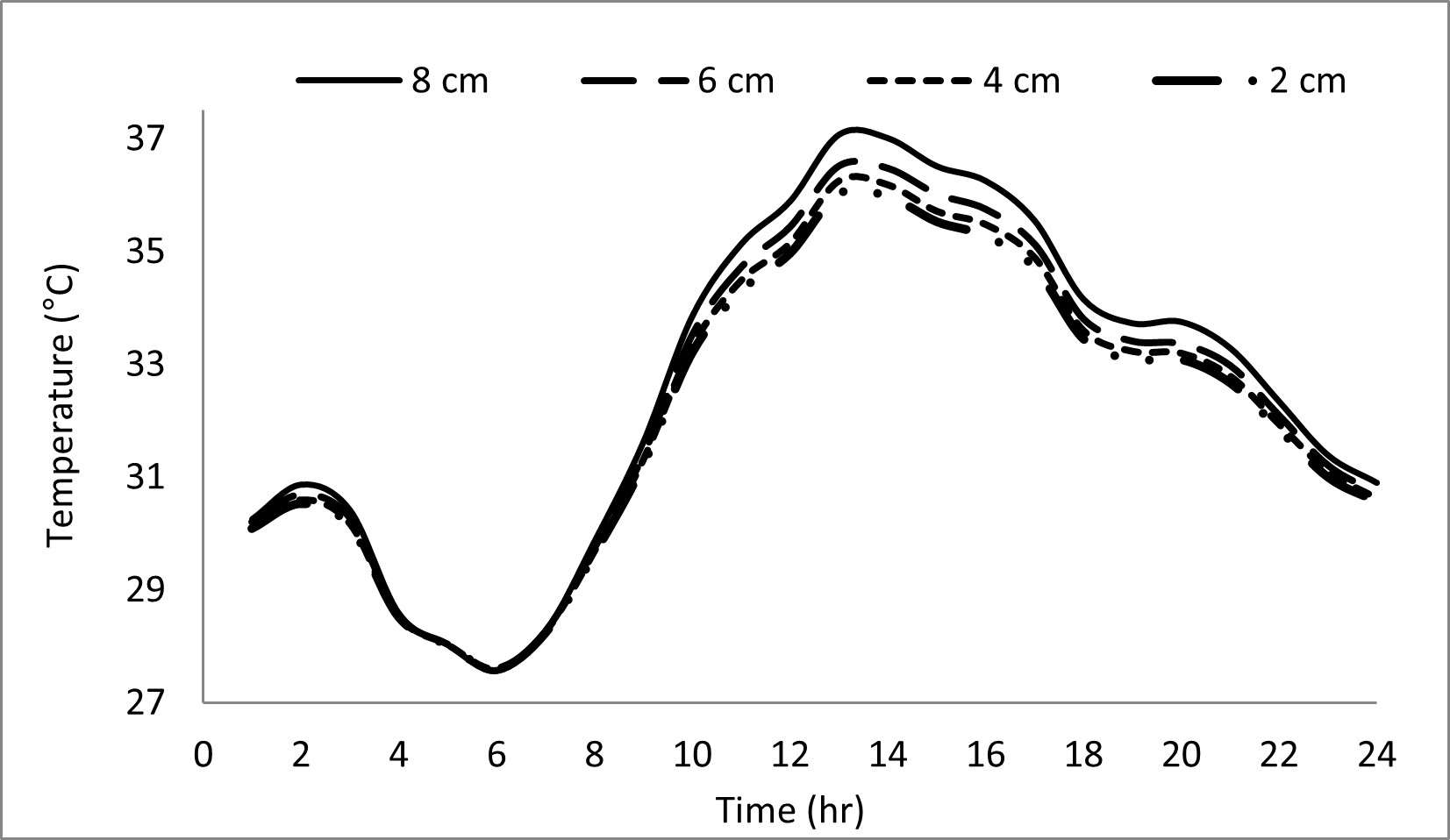}
    \caption{The graph of temperature changes in one day for different types of buried pipes }
\end{figure}

Figure 13 displays the results of an investigation into the impact of the buried pipe diameter parameter using four sizes of 2, 4, 6, and 8 cm. The temperature nearly did not change from 3 to 8 o'clock, as shown in Figure 13, which may be attributed to the building's temperature being similar to that of the soil. However, throughout the remaining hours, as the diameter increased from 2 to 8 cm, the temperature increased. According to equation 36, the reason why ventilation effectiveness falls as the room size grows is that as the diameter grows, so does the heat transfer coefficient.

\begin{figure}[h!]
    \centering   \includegraphics[width=0.55\textwidth]{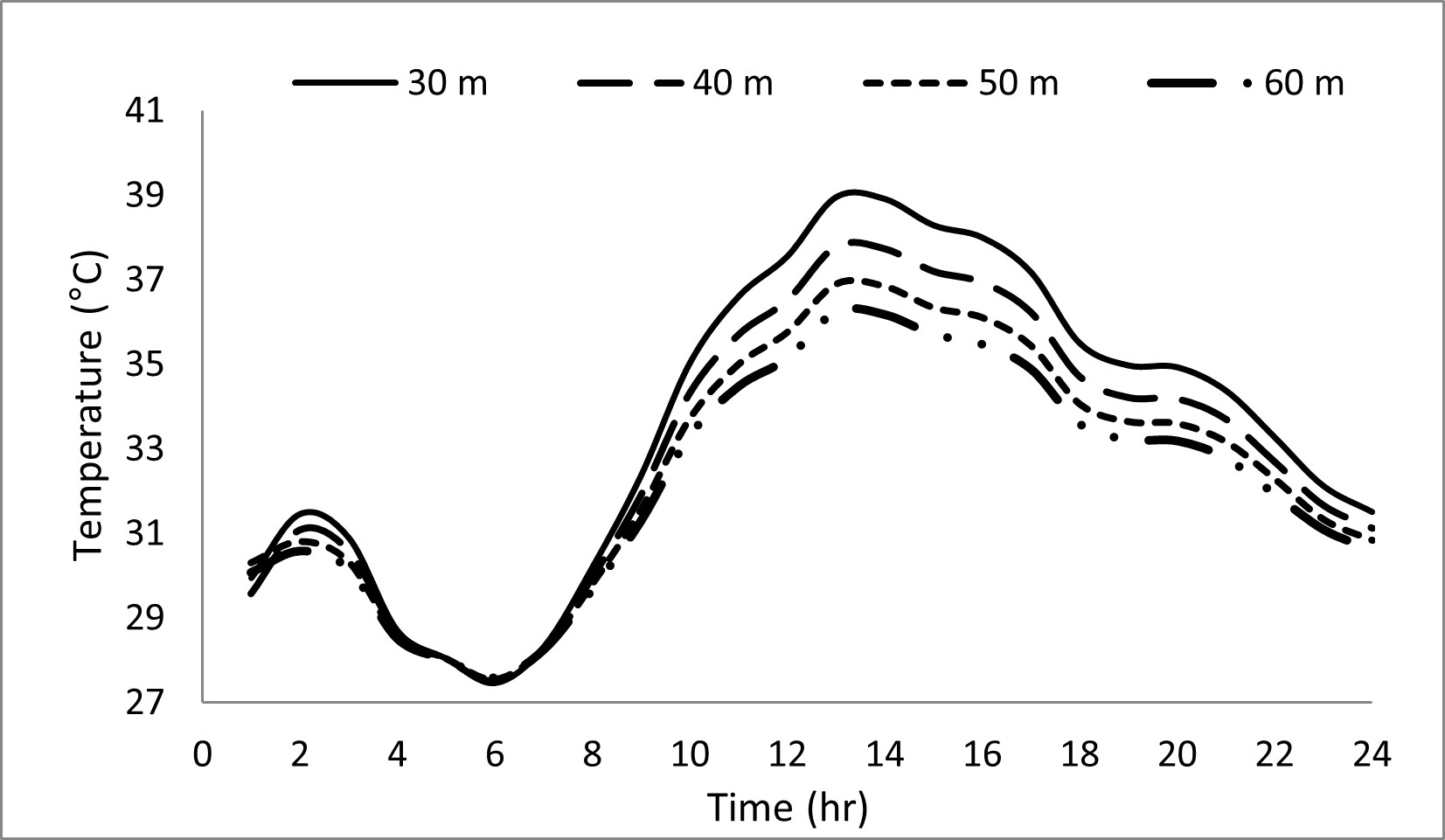}
    \caption{The graph of temperature changes in one day for different lengths of buried pipe}
\end{figure}

Four lengths of 30, 40, 50, and 60 meters were examined to survey the impact of the buried pipe length parameter; the result is shown in Figure 14. The temperature has not changed much between 4 and 7 o'clock, as shown in Figure 14, and this may be attributed to the building's temperature being near to that of the earth. However, throughout the remaining hours, the length grows from 30 meters to 60 meters. Reduced room temperature and improved ventilation efficiency. The longer they spend in the soil and the resulting colder air are the causes of it\cite{weeratunge2021feasibility} Four flow rates were examined to determine the impact of the air input flow parameter in the buried pipe, and the results are shown in Figure 15. As shown in Figure, the temperature has hardly changed between the hours of 4 and 7 o'clock, and this can be attributed to the fact that the building's diameter and length parameters are similar to the soil temperature. However, throughout the remaining hours, the flow rate increased. Reduced room temperature and improved ventilation efficiency. The cause of this may be explained by equation 36, which states that as the flow rate rises, so does the heat transfer coefficient.

\vspace{25pt}
\begin{figure}[h!]
    \begin{minipage}[c]{0.46\textwidth}
         \centering
    \includegraphics[width=\textwidth]{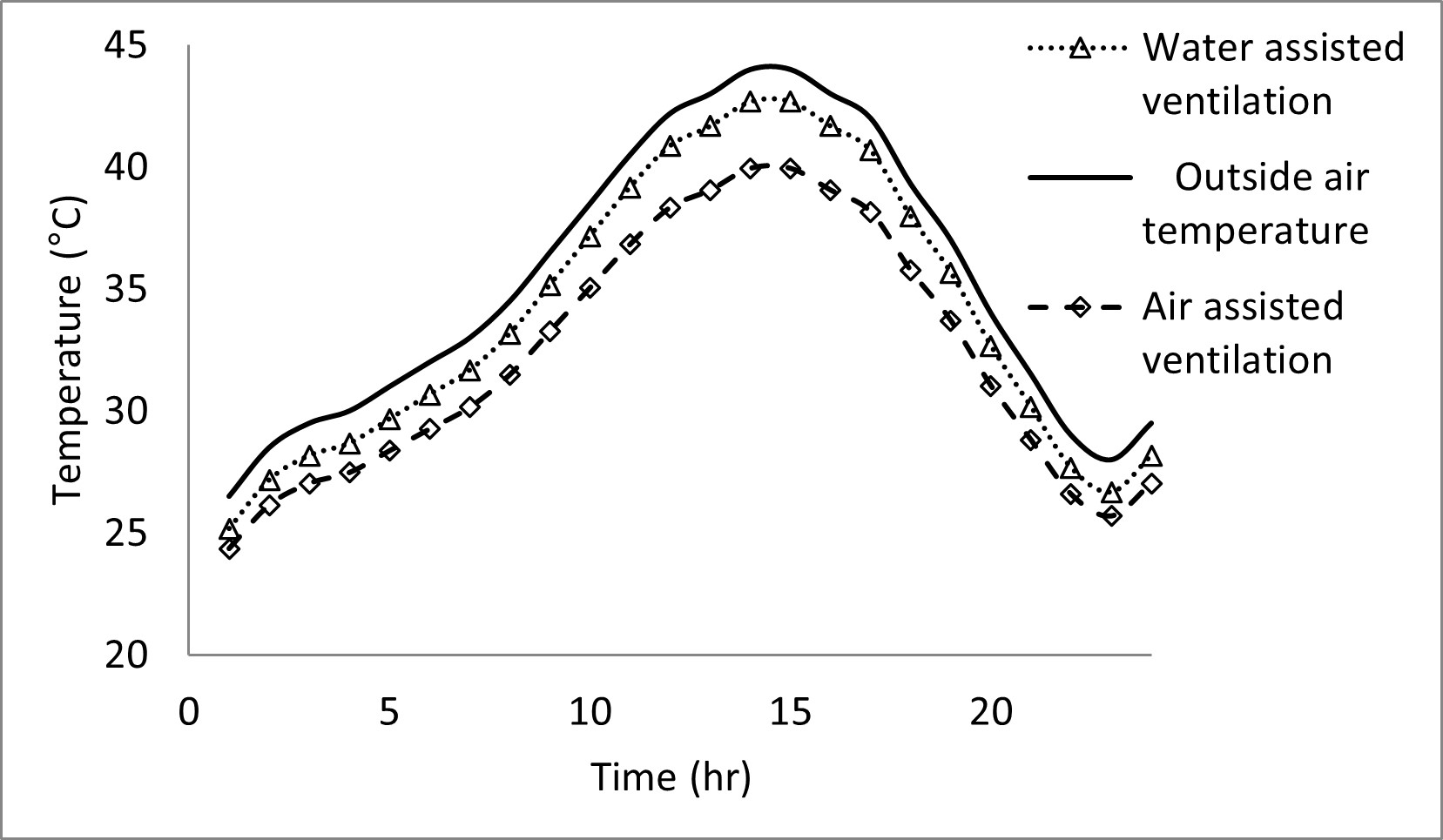}
     \caption{Temperature changes chart in one day for different fluids}
  \end{minipage}
  \hfill
  \begin{minipage}[c]{0.46\textwidth}
     \centering
  \includegraphics[width=\textwidth]{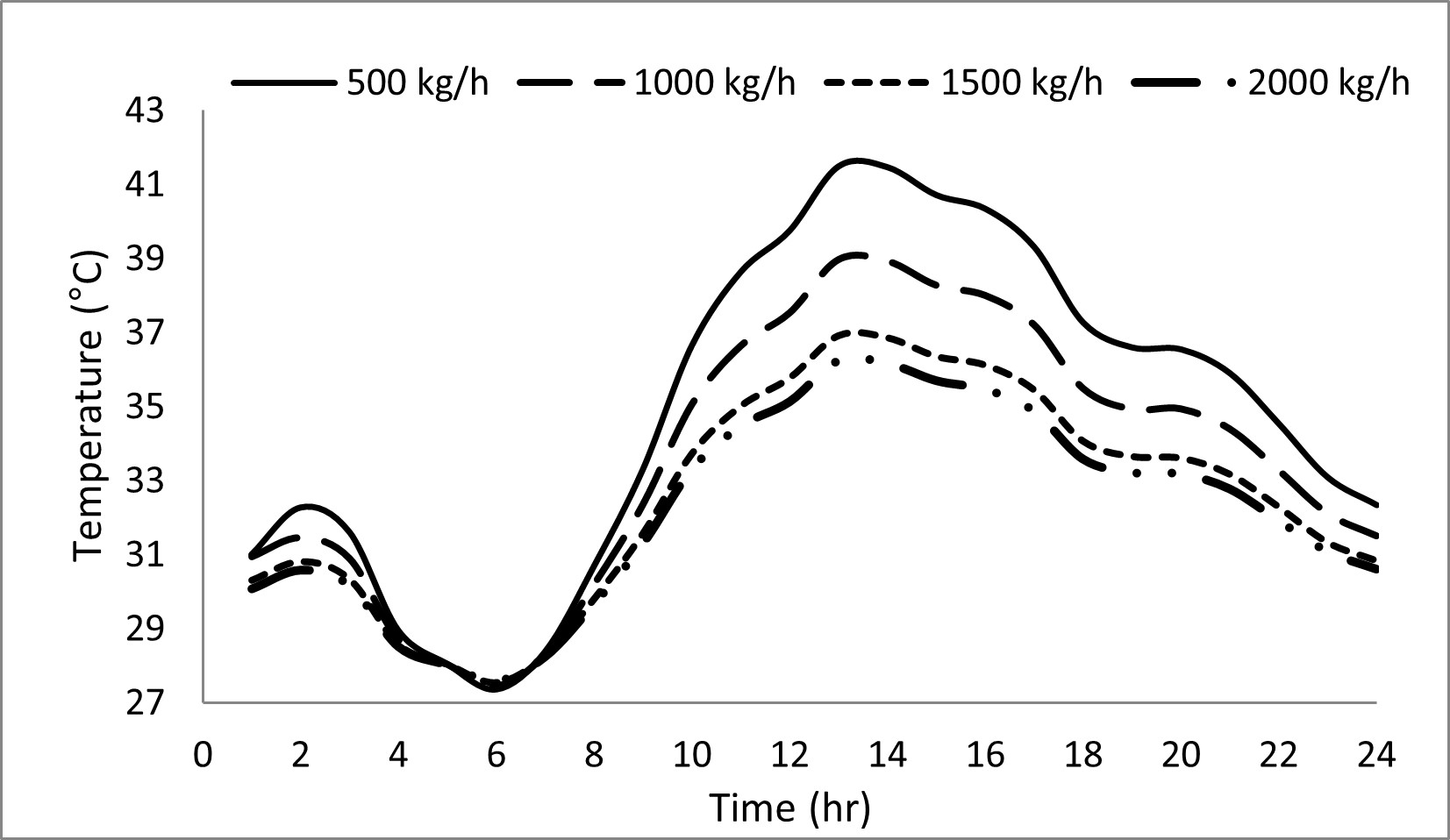}
    \caption{Temperature changes chart in one day for different flow rates of the buried pipe}
  \end{minipage}
\end{figure}

Two kinds of air and water fluid have been taken into consideration to compare the ventilation system in order to explore the effect of the fluid parameter in the buried pipe. The outcome is shown in Figure 16. The building ventilated with water fluid has a greater temperature than the building ventilated with air-fluid, as can be observed. This demonstrates that a ventilation system using air is more effective than one using water.

\subsection{Greenhouse Structure Results}
Figures 17 to 24 show the result of the examination of this heating and cooling system in four distinct cities: Tehran, Tabriz, Abadan, and Rasht for a greenhouse structure.
\vspace{25pt}
\begin{figure}[h]
    \begin{minipage}[c]{0.46\textwidth}
         \centering
    \includegraphics[width=\textwidth]{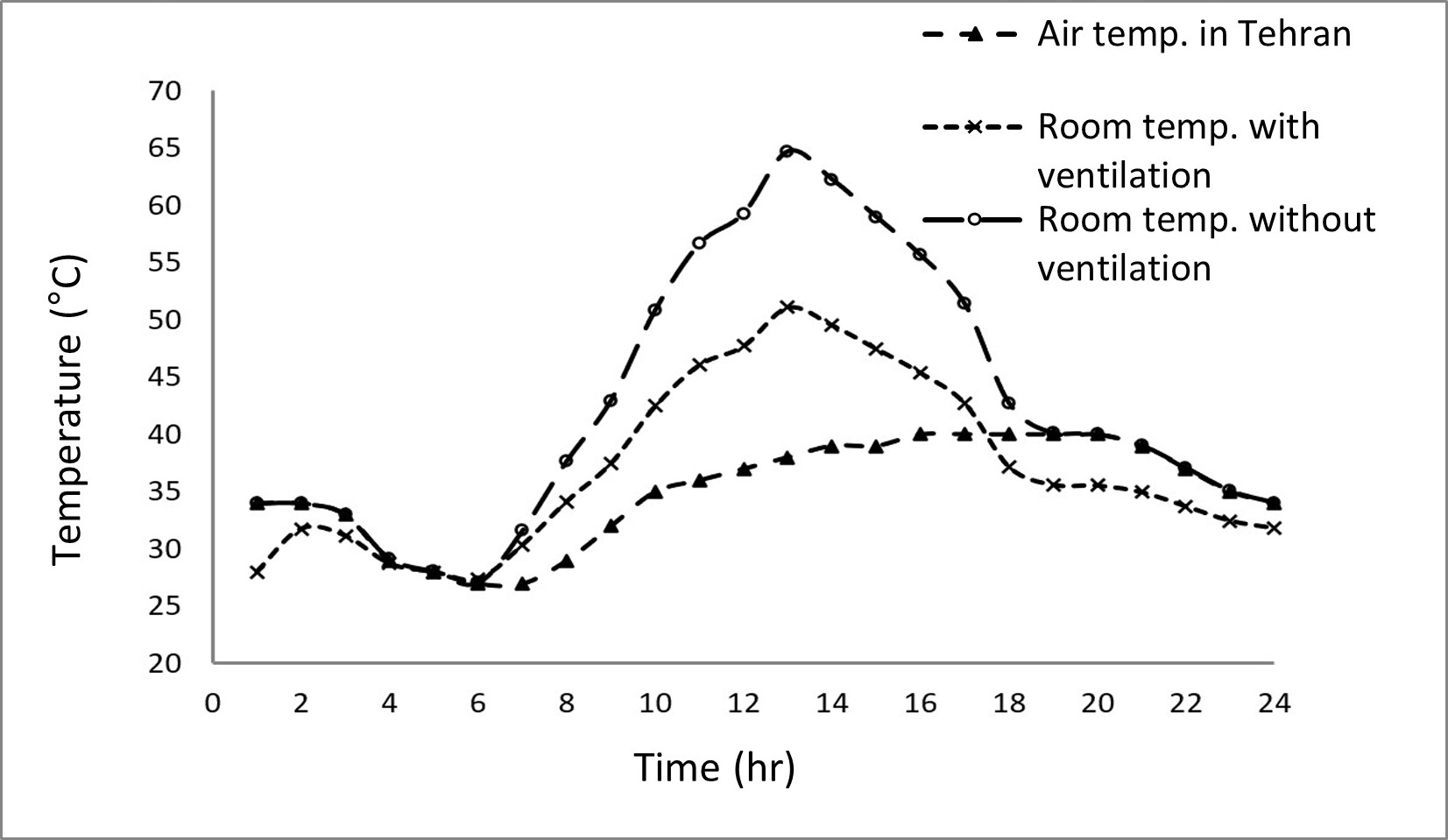}
     \caption{Temperature changes diagram at different times of the hottest day in Tehran}
  \end{minipage}
  \hfill
  \begin{minipage}[c]{0.46\textwidth}
     \centering
  \includegraphics[width=\textwidth]{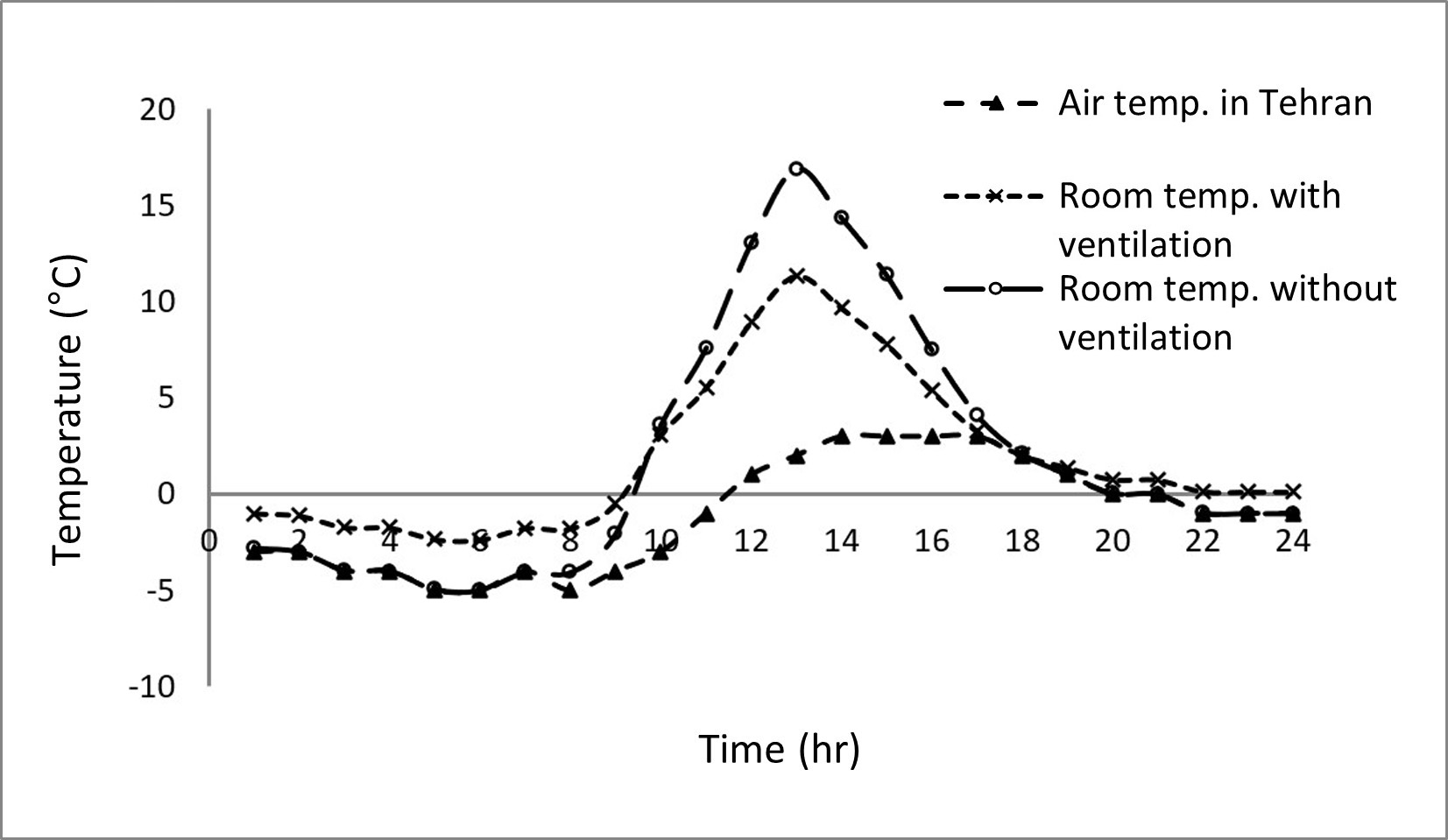}
    \caption{Temperature changes chart at different times of the coldest day in Tehran}
  \end{minipage}
\end{figure}

\begin{figure}[h]
    \begin{minipage}[c]{0.46\textwidth}
         \centering
    \includegraphics[width=\textwidth]{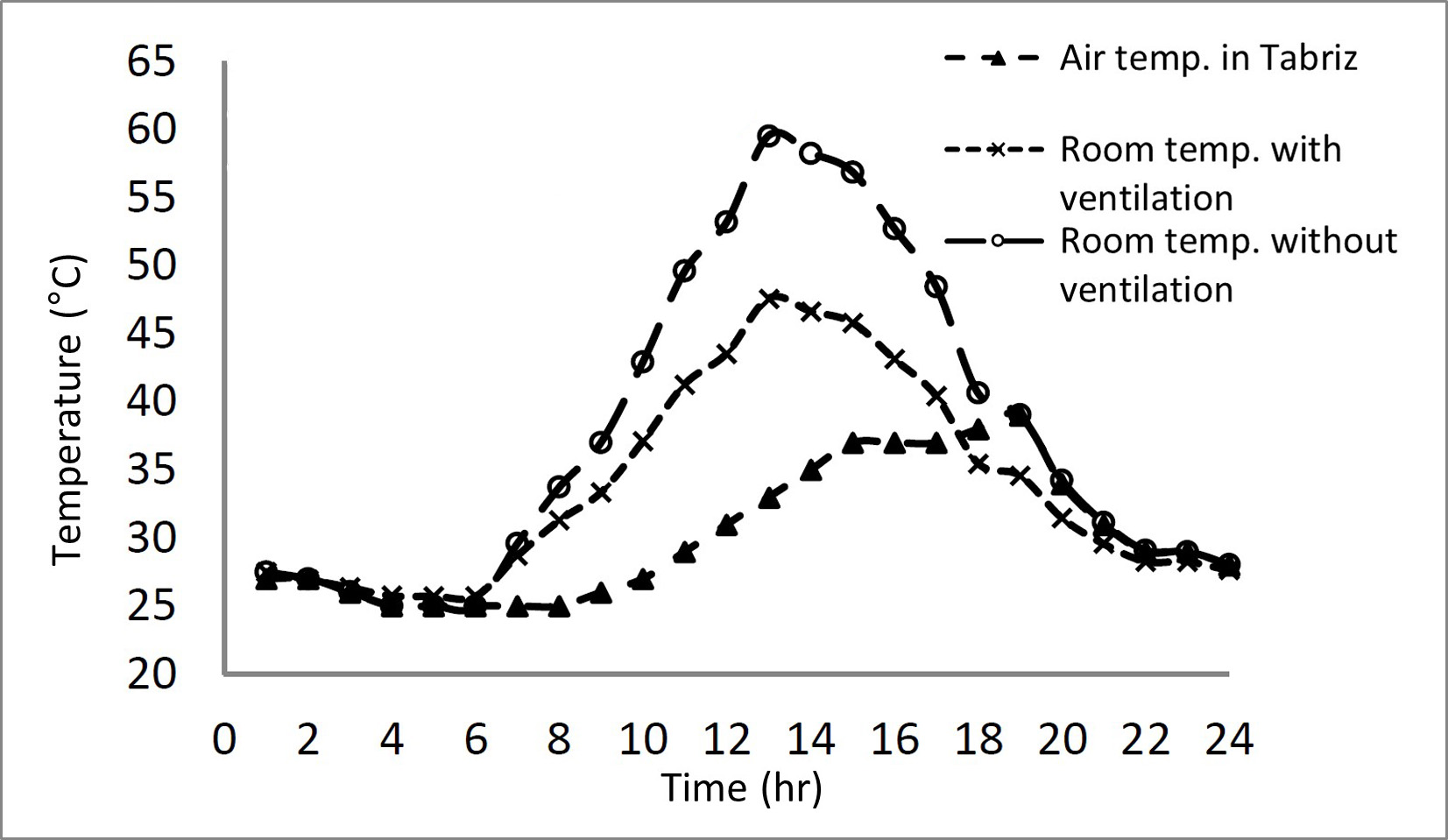}
     \caption{Temperature changes diagram at different times of the hottest day in Tabriz}
  \end{minipage}
  \hfill
  \begin{minipage}[c]{0.46\textwidth}
     \centering
  \includegraphics[width=\textwidth]{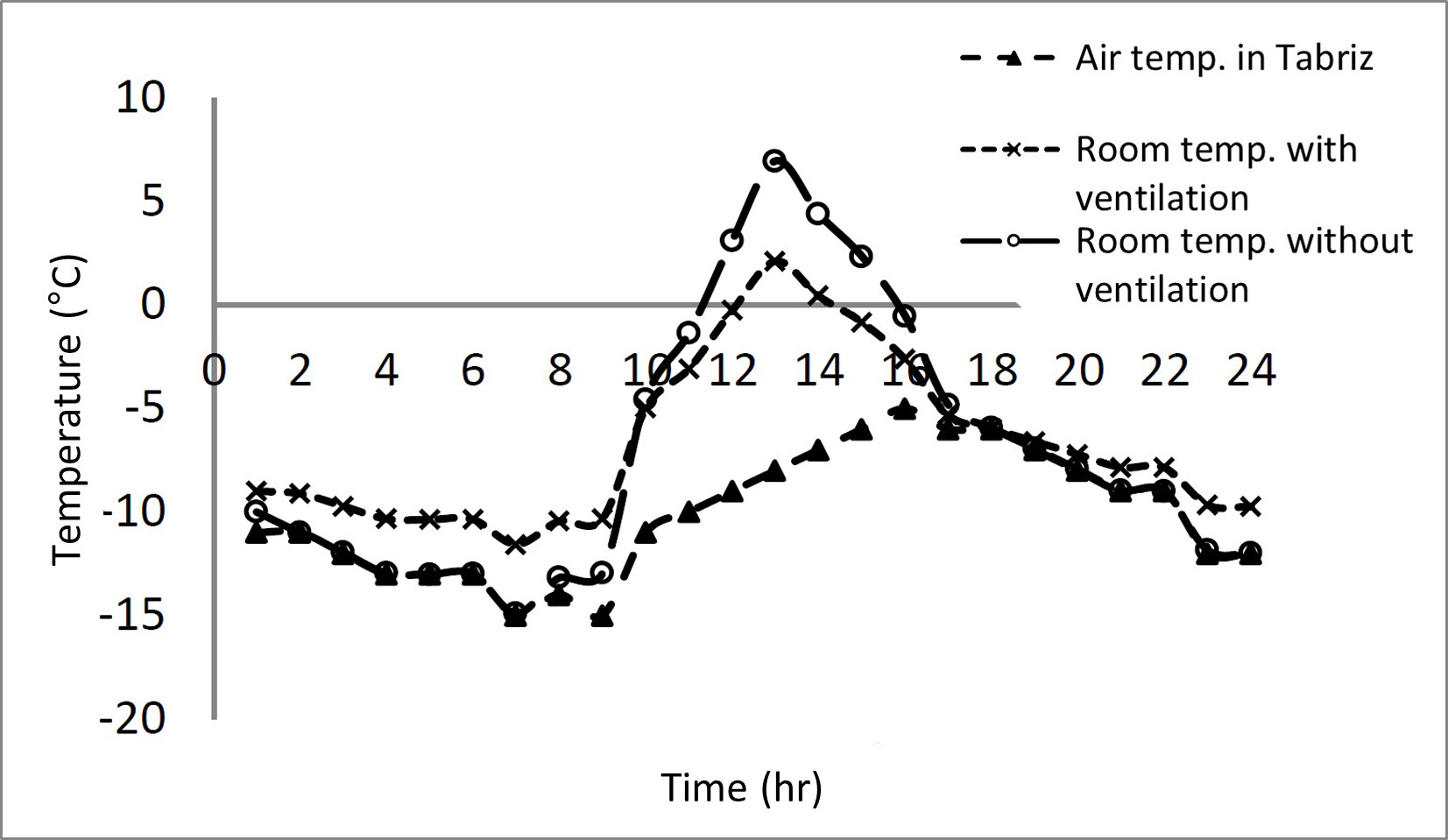}
    \caption{Temperature changes chart at different times of the coldest day in Tabriz}
  \end{minipage}

    \begin{minipage}[c]{0.46\textwidth}
         \centering
    \includegraphics[width=\textwidth]{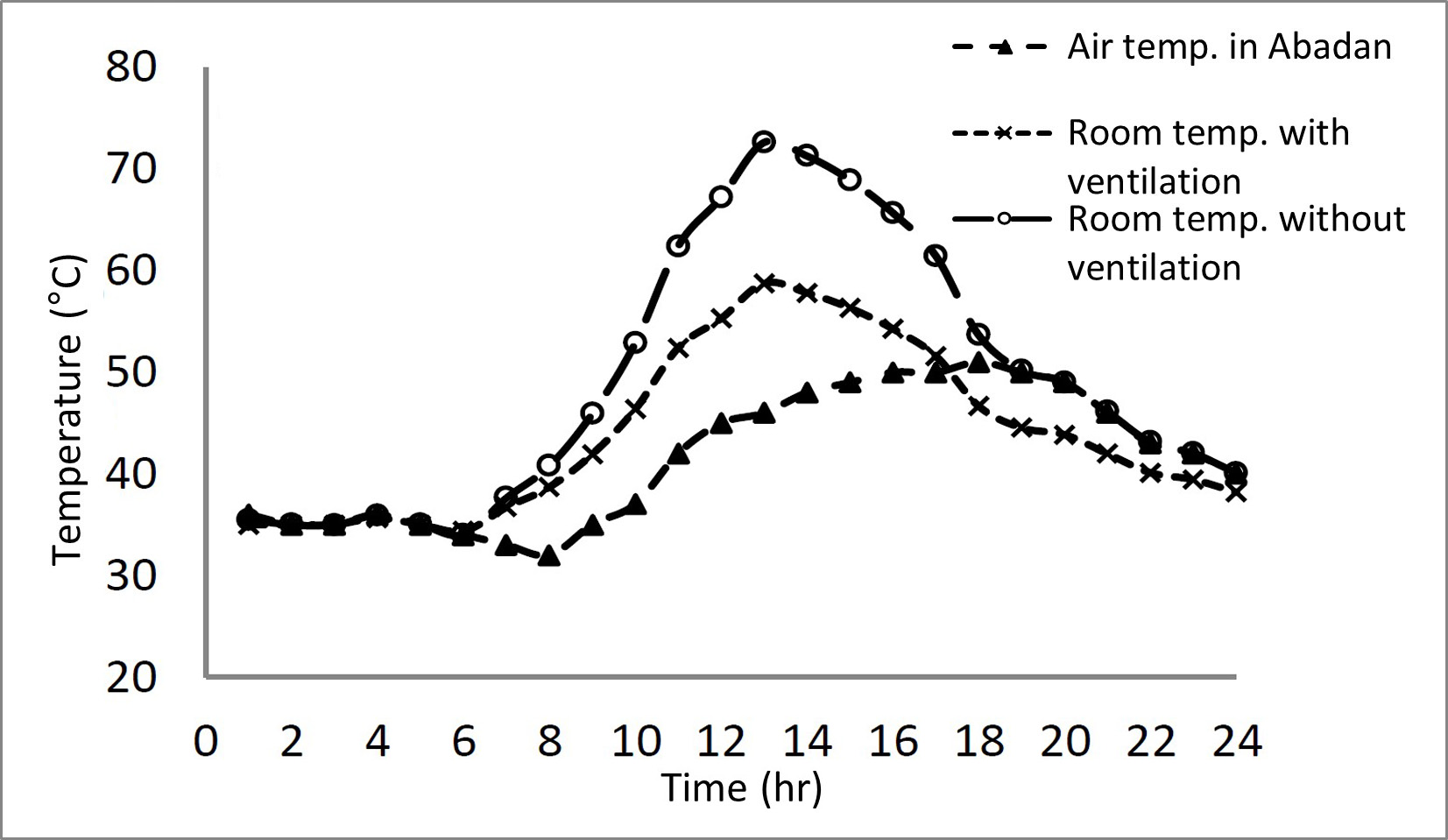}
     \caption{Temperature changes diagram at different times of the hottest day in Abadan}
  \end{minipage}
  \hfill
  \begin{minipage}[c]{0.46\textwidth}
     \centering
  \includegraphics[width=\textwidth]{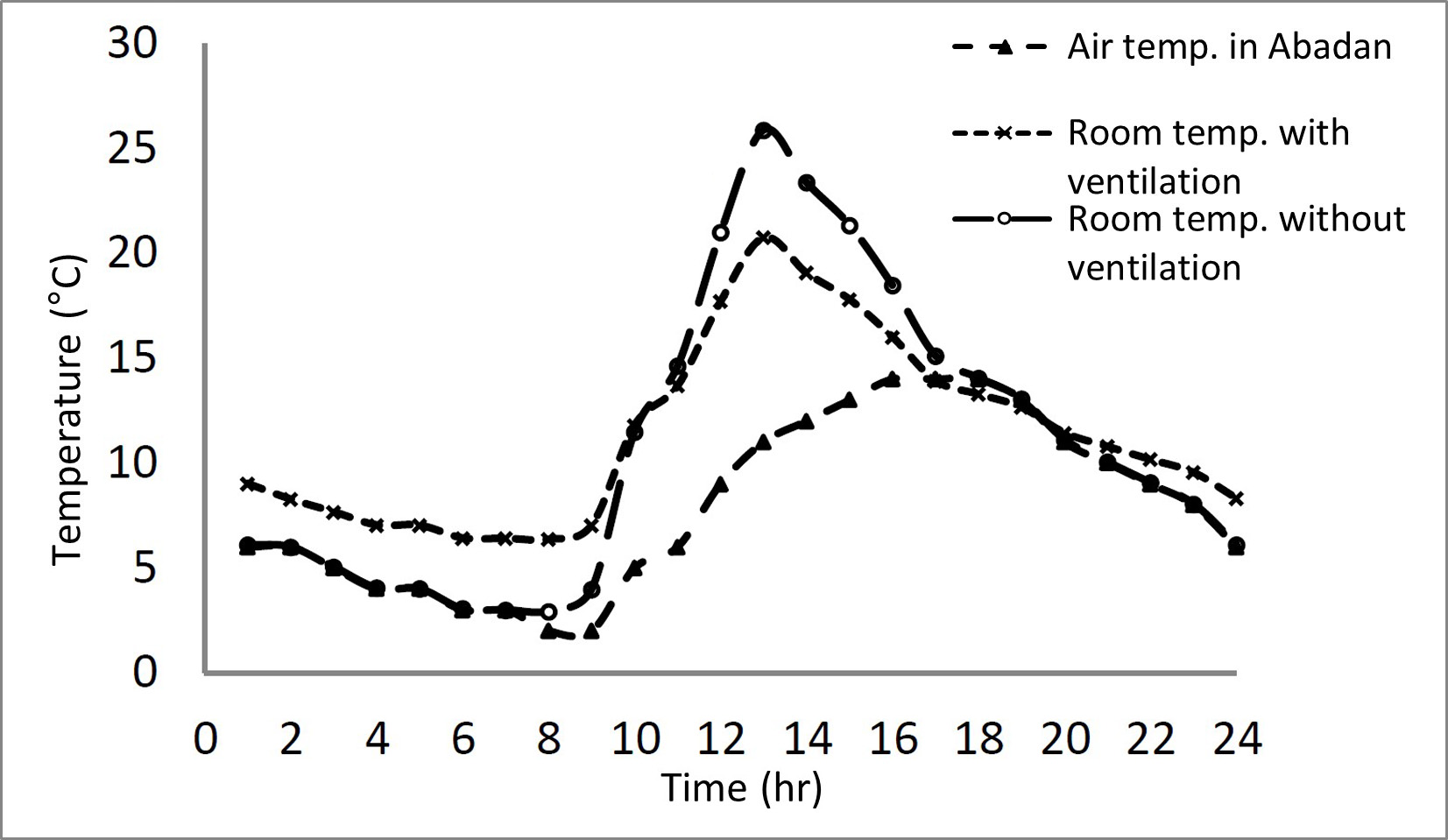}
    \caption{Temperature changes chart at different times of the coldest day in Abadan}
  \end{minipage}
\end{figure}

\begin{figure}[h!]
    \begin{minipage}[c]{0.46\textwidth}
         \centering
    \includegraphics[width=\textwidth]{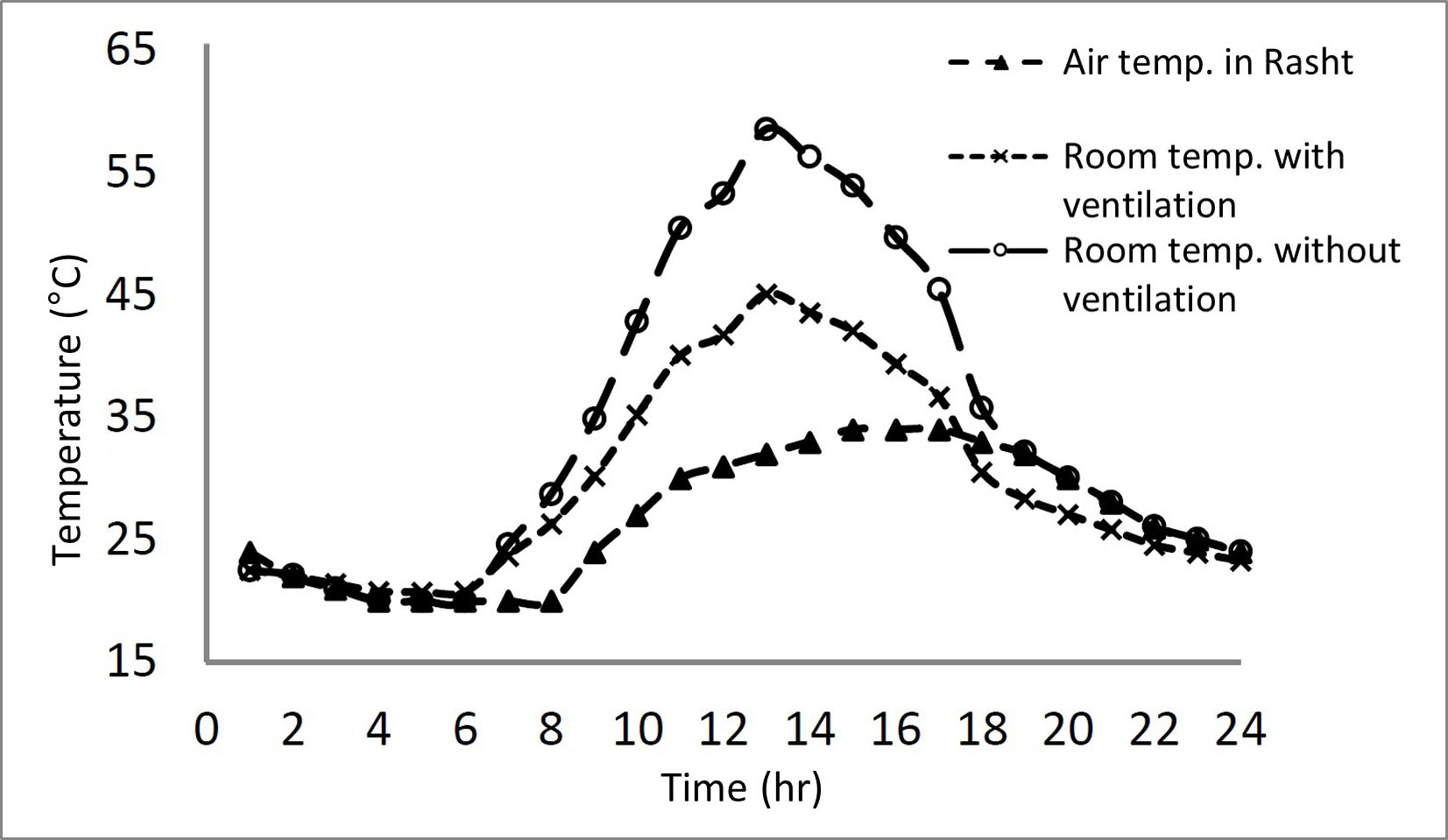}
     \caption{Temperature changes diagram at different times of the hottest day in Rasht}
  \end{minipage}
  \hfill
  \begin{minipage}[c]{0.46\textwidth}
     \centering
  \includegraphics[width=\textwidth]{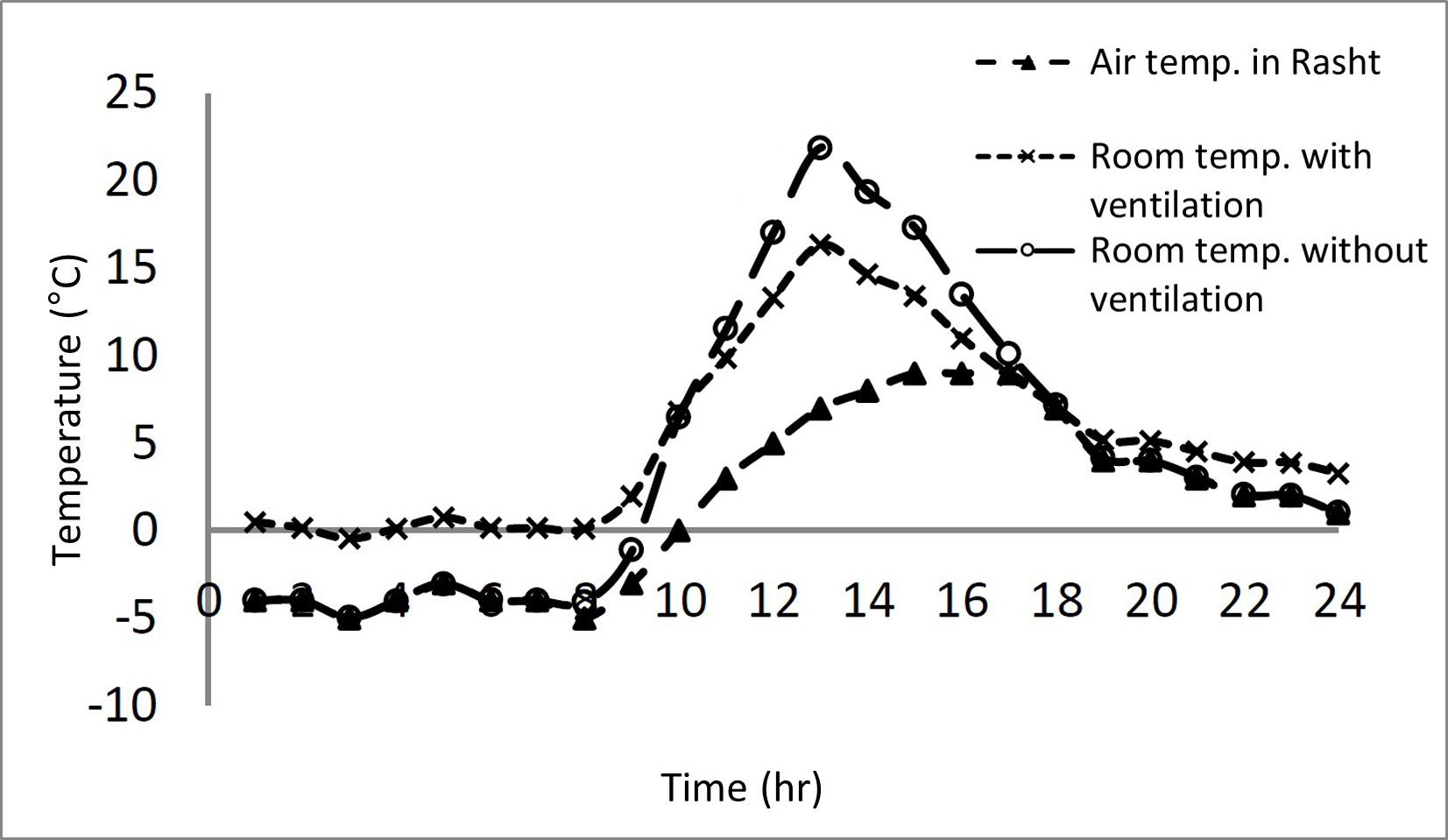}
    \caption{Temperature changes chart at different times of the coldest day in Rasht}
  \end{minipage}
\end{figure}

\vspace{50pt}
\section{Expenses}
The implementation and maintenance expenses of this study should be looked into and contrasted with those of other approaches in order to evaluate its economic efficacy. In this study, a cooling system is employed in the summer and a soil heating system in the winter. The air temperature may be utilized in winter because the soil temperature is greater than the air temperature, and in summer because the soil temperature is lower. If this system uses air to operate, it should need a blower to move the air within the system and several meters of tubing to be buried one meter into the earth\cite{yumrutas2012modeling}. If this system uses water, it will need a pump to circulate the water within the pipe, several meters of pipe to be buried in the ground, as well as a fan coil to circulate the air inside the building using either cold water or compressed air. It must be heated with the aid of soil. As can be seen, using this mechanism with the assistance of air is more effective than using water. The cost of the water heating and cooling system is more than that of the air heating and cooling system, as was already established. Split is an alternative technique that may be utilized for both heating and cooling. Table 4 displays the expenses for the split and condenser. 

\begin{table*}[h!]
\centering
  \caption{Pipe costs}
  \label{tab:commands}
  \begin{tabular}{ccl}
    \toprule
    Description & Price(\$) & Pipe\\
    \midrule
    \texttt Low corrosion and low heat transfer coefficient  & 2 & Price per meter of polycan pipe  \\
    \texttt Corrosive and high heat transfer coefficient  & 2.75 & Price per meter of pipe with galvanized sheet  \\
    \bottomrule
  \end{tabular}
\end{table*}

Two varieties of Polycab pipe, which have a lower heat transfer coefficient, and galvanized sheet pipe, which have a greater heat transfer coefficient, are listed in Table 2 along with the pipe pricing. Due to its greater ability to transport heat, this pipe has a galvanized layer covering it. The associated cost will be about 165 dollars if a pipe size of 60 meters is chosen. 

\begin{table}[h!]
\centering
  \caption{Air blower costs \cite{kizikoglu2015thermo}}
  \label{tab:commands}
  \begin{tabular}{c c}
    \toprule
    Price(\$) & Blower flow rate (cubic meters per hour)\\
    \midrule
    \texttt 80 & 0.15 \\
    \texttt 140 & 0.30 \\
    \texttt 180 & 0.40 \\
    \texttt 230 & 0.50 \\
    \bottomrule
  \end{tabular}
\end{table}

According to two tables 2 and 3, the cost of heating and cooling system by helping soil is obtained from 250 dollars to 350 dollars, which is lower according to the costs due to Split, which is given in Table 4. So it can be said that this system is almost affordable. 

\begin{table*}[h!]
\centering
  \caption{Costs related to Split}
  \label{tab:commands}
  \begin{tabular}{ccc}
    \toprule
    Price per meter of copper pipe and insulation(\$)  & Price(\$) & Split\\
    \midrule
    \texttt 8 & 600 & 9000(BTU)  \\
    \texttt 8 & 700 & 12000(BTU)  \\
    \bottomrule
  \end{tabular}
\end{table*}

Additionally, Table 5 displays the price of power used by each of the systems indicated. Each kilowatt hour of power used is estimated to cost 0.02 Dollars. As can be seen, the cost of power for a 30-day period with a 24-hour on-time for the air blower, pump, fan coil, and split is around 2.8 dollars, 3.8 dollars, 9.5 dollars, and 24 dollars, respectively. It should be noted that, as was already said, the fan coil and pump are used for ventilation using water and dirt. Therefore, it may be stated that the ventilation system that uses dirt and air has the lowest cost. 

\begin{table*}[h!]
\centering
  \caption{Costs related to electricity consumption}
  \label{tab:commands}
  \begin{tabular}{cccc}
    \toprule
    Device Name  & amount of electric current (amps) & power (watts) & One month's electricity consumption (\$) \\
    \midrule
    \texttt Air Blower & 0.9 & 198 & 2.8512  \\
    \texttt Pump & 1.2 & 264 & 3.8016 \\
    \texttt Fan Coil & 3 & 660 & 9.504 \\
    \texttt Split & 7.6 & 1672 & 24.0768 \\
    \bottomrule
  \end{tabular}
\end{table*}

\section{Conclusion}

In this paper, the following results were provided: 

- In Tehran, the soil ventilation system reduces the maximum temperature by 10 degrees in the summer and raises the lowest temperature of the residential building by 4 degrees in the winter. 

- In Tabriz, a soil ventilation system raises the minimum temperature of residential buildings by 6 degrees in the winter and lowers them by 10 degrees in the summer. 

- Abadan City's ventilation system uses dirt to raise the minimum temperature of residential buildings by 6 degrees in the winter and lower them by 12 degrees in the summer. 

- In Rasht City, the soil ventilation system raises the minimum temperature of the residential building by 7 degrees in the winter and lowers it by 11 degrees in the summer.

- In Tehran, the soil ventilation system reduces the maximum temperature by 14 degrees in the summer and raises the lowest temperature of the greenhouse by 3 degrees in the winter. 

- In Tabriz, a soil ventilation system raises the minimum temperature of the greenhouse by 3 degrees in the winter and lowers them by 13 degrees in the summer. 

- Abadan City's ventilation system uses dirt to raise the minimum temperature of the greenhouse by 4 degrees in the winter and lower them by 15 degrees in the summer. 

- In Rasht City, the soil ventilation system raises the minimum temperature of the greenhouse by 5 degrees in the winter and lowers it by 13 degrees in the summer.

- The ventilation system performed better in the summer than in the winter, it may be claimed. 

- The temperature hardly changed when the diameter of the buried pipe was changed from 3 to 8 o'clock; this can be attributed to the fact that the temperature of the building is similar to that of the soil, but during the other hours, when the diameter was increased from 2 to 8 cm, the temperature of the room increased. It rises throughout the summer, and ventilation effectiveness falls. 

- When the buried pipe's length was changed from 4 to 7 o'clock, the temperature almost did not change, and the reason for this can be attributed to the building's temperature being close to the soil's temperature. However, when the length was increased from 30 to 60 meters during the remaining hours, the temperature increased. In the summer, the space is smaller and ventilation is more effective. The longer they spend in the soil and the resulting colder air are the causes of it. 

- From 4 to 7 o'clock, the airflow in the buried pipe was changed, but the temperature did not change. This can be attributed to the fact that the building's diameter and length parameters are close to the soil temperature, but during the other hours, as the flow rate is increased, the room temperature in the summer is lowered, and ventilation effectiveness is raised. 

- The building's temperature is greater when it is ventilated with water than it is ventilated with air. This demonstrates that using air to ventilate a space is more effective than using water. 

- The ventilation system that uses soil and air as ventilation has the lowest cost.
\bibliographystyle{unsrt}  
\bibliography{references}

\end{document}